\documentstyle[12pt,aaspp4]{article}
\begin{document}

\title{The {\it COBE} Diffuse Infrared Background Experiment Search for the
Cosmic Infrared Background: III. Separation of Galactic Emission from the
Infrared Sky Brightness} 

\author{
R. G. Arendt,\altaffilmark{1,2}
N. Odegard,\altaffilmark{2}
J. L. Weiland,\altaffilmark{2}
T. J. Sodroski,\altaffilmark{3}
M. G. Hauser,\altaffilmark{4}
E. Dwek,\altaffilmark{5}
T. Kelsall,\altaffilmark{5}
S. H. Moseley,\altaffilmark{5}
R. F. Silverberg,\altaffilmark{5}
D. Leisawitz,\altaffilmark{6}
K. Mitchell,\altaffilmark{7}
W. T. Reach,\altaffilmark{8}\\
E. L. Wright\altaffilmark{9}
}
\altaffiltext{1}{email: arendt@stars.gsfc.nasa.gov}
\altaffiltext{2}{Raytheon STX, Code 685, NASA GSFC, Greenbelt, MD 20771} 
\altaffiltext{3}{Space Applications Corp., Code 685, NASA GSFC, Greenbelt, MD
20771} 
\altaffiltext{4}{Space Telescope Science Institute, 3700 San Martin Drive, 
Baltimore MD, 21218}
\altaffiltext{5}{Code 685, NASA GSFC, Greenbelt, MD 20771}
\altaffiltext{6}{Code 631, NASA GSFC, Greenbelt, MD 20771}
\altaffiltext{7}{General Sciences Corp., Code 920.2, NASA GSFC, Greenbelt, MD
20771} 
\altaffiltext{8}{IPAC, Caltech, MS 100-22, Pasadena, CA 91125}
\altaffiltext{9}{UCLA, Department of Astronomy, Los Angeles, CA 90024-1562}

\begin{center}
submitted to The Astrophysical Journal\\
received: 8 January 1998, accepted: 12 May 1998
\end{center}

\begin{abstract}

The Cosmic Infrared Background (CIB) is hidden behind veils of foreground 
emission from our own solar system and Galaxy. This paper describes 
procedures for removing the Galactic IR emission from the 1.25 -- 240 $\micron$
{\it COBE} DIRBE maps as steps toward the ultimate goal of detecting the CIB. 
The Galactic emission models are carefully chosen and constructed so
that the isotropic CIB is completely retained in the residual sky maps. 
We start with DIRBE data from which the scattered light and thermal emission
of the interplanetary dust (IPD) cloud have already been removed. 
Locations affected by the emission from bright compact and stellar 
sources are excluded from the analysis. The unresolved emission of faint stars
at near- and mid-IR wavelengths is represented by a model based on 
Galactic source counts. The 100 $\micron$
DIRBE observations are used as the spatial template for the interstellar
medium (ISM) emission at
high latitudes. Correlation of the 100 $\micron$ data with H I column density
allows us to isolate the component of the observed emission that is associated
with the ISM. Limits are established on the far-IR emissivity of the diffuse
ionized medium, which indicate a lower emissivity per H nucleus than in the 
neutral medium.
At 240 $\micron$, we find that adding a second spatial template
to the ISM model can greatly improve the accuracy of the model at low
latitudes. The crucial product of this analysis is a set of all-sky IR maps
from which the Galactic (and IPD) emission has been removed. We discuss
systematic uncertainties and potential errors in the foreground subtraction
process that may have an impact on studies seeking to detect the CIB in the
residual maps. 

\end{abstract}

\keywords{diffuse radiation --- Galaxy: general --- infrared: ISM: continuum 
--- infrared: stars --- ISM: general}

\section{Introduction}

The primary scientific goal of the Diffuse Infrared Background Experiment
(DIRBE) aboard the {\it Cosmic Background Explorer (COBE)} spacecraft
is the measurement of the cosmic infrared background (CIB) at 
wavelengths from 1.25 to 240 $\micron$. This radiation is the cumulative
emission of pregalactic sources, protogalaxies, and evolving galaxies, as
well as emission from more exotic processes not common in the local universe
(e.g. Bond, Carr, \& Hogan 1986). These
sources of the CIB may be found anywhere from the earliest epoch after
radiation and matter were decoupled to the present day. 
The contribution to the CIB from galaxies will
be composed of stellar emission from distant galaxies that is redshifted by the
cosmological expansion from intrinsically shorter wavelengths, as well as the
direct IR emission from stars and dust within galaxies at all distances. 
The constraints that 
DIRBE places on the CIB are discussed by Hauser et al. (1998, hereafter Paper 
I) and Dwek et al. (1998, hereafter Paper IV).

However, in order to detect the CIB, we first need to remove the strong
contributions of foreground emission arising within our own solar system and
Galaxy. The IR foreground from within our solar system originates
from the interplanetary dust (IPD) cloud. The modeling 
and removal of the scattering and emission from the IPD for the entire
cold-mission DIRBE data set are reported by Kelsall et al. (1998, hereafter 
Paper II). Following
the removal of this foreground, we attack the next layer of foreground
emission by modeling and subtracting the Galactic IR emission. 

The near-IR (1.25 -- 4.9 $\micron$) emission of the Galaxy is dominated by
starlight. Some bright stars and other compact sources are resolved as point
sources in the DIRBE data. Most stars blend into a diffuse background, showing
a disk and bulge very similar in appearance to many edge-on spiral galaxies.
Extinction effects are clearly present at the shortest near-IR wavelengths as
a visible dark lane in the inner Galaxy. Papers which have 
previously examined DIRBE observations of the stellar disk and bulge of the 
Galaxy are: 
Weiland et al. (1994), 
Arendt et al. (1994), 
Freudenreich et al. (1994), 
Dwek et al. (1995), 
Calbet et al. (1996), 
Freudenreich (1996), 
Binney, Gerhard, \& Spergel (1997), 
Bissantz et al. (1997), 
Porcel, C., Battaner, E., \& Jim\'enez-Vicente, J. (1997), and 
Fux (1997). 
The mid- and
far-IR (12 -- 100 and 140 -- 240 $\micron$) emission is dominated by thermal
emission from dust in the diffuse ISM and in more compact star-forming
regions. Previously published studies of the Galactic ISM based on DIRBE data
include: 
Arendt et al. (1994), 
Freudenreich et al. (1994), 
Sodroski et al. (1994, 1995), 
Bernard et al. (1994), 
Boulanger et al. (1996), 
Dwek et al. (1997), 
Sodroski et al. (1997), 
Davies et al. (1997), and
Lagache et al. (1998).

This paper details the development of models of the Galactic IR emission.
The primary intended use for the models is to permit an accurate
measurement of an extragalactic IR background. The characterization of the
sources that give rise to the Galactic emission is an important secondary
result of the process. An overview of the procedures used for modeling the
Galactic foreground is given in Section 2. Section 3 describes the DIRBE data
set and its preparation, particularly the removal of the IPD scattering and 
emission (Paper II). Section 4 describes the modeling and removal of
starlight from the 1.25 to 25 $\micron$ measurements. The following section
(\S5) describes in detail the modeling and removal of the ISM emission from
DIRBE data. 
Additional investigation into the far-IR emissivity of the diffuse ionized ISM
is contained in the Appendix.
The accuracy of the removal of the Galactic emission is a major
limitation to the detection of the CIB, and consequently Section 6 discusses
estimates of the uncertainties of the data and procedures. A
brief discussion of the implications of this modeling for Galactic properties
is contained in Section 7. A more detailed analysis of the Galactic ISM as
revealed by this work has been reported by Dwek et al. (1997). Finally, a
brief summary of results is given in Section 8. Companion papers (Papers I \& 
IV) contain analyses of the residual emission, and its implications for
the detection of the CIB. 

\section{Methods}

We will assume that at any time, $t$, 
the total intensity observed at a given wavelength and a
specified Galactic longitude and latitude, $I_{obs}(l,b,\lambda,t)$, can be 
represented by:
\begin{equation}
I_{obs}(l,b,\lambda,t) = Z(l,b,\lambda,t) + G_S(l,b,\lambda) + 
G_I(l,b,\lambda) + I_0(l,b,\lambda)
\end{equation}
where $Z(l,b,\lambda,t)$ is the contribution from the interplanetary
dust cloud, $G_S(l,b,\lambda)$ is the contribution from stars (resolved and 
unresolved) and other compact sources within the Galaxy, $G_I(l,b,\lambda)$ is
the contribution from the interstellar medium, and $I_0(l,b,\lambda)$ is
residual intensity obtained by subtracting the former three components from
the total observed intensity. The procedures we apply are designed to ensure 
that the CIB is fully contained within this residual emission, i.e., that 
no CIB component is inadvertently included in the models of the IPD cloud 
or Galactic emission components. 

The contribution of the IPD to the infrared sky, $Z(l,b,\lambda,t)$, has been
calculated in Paper II. The IPD cloud is modeled using a
geometric kernel similar to that developed to represent IPD emission in the
{\it IRAS} data (Wheelock et al. 1994). The model ensures the preservation of
the CIB intensity by fitting the time variation of the IPD emission rather
than the intensity directly. The model does not contain any explicit isotropic
component which would produce no time variation. Details of this model are
presented in Paper II.

In the DIRBE data, the stellar emission component of the Galaxy,
$G_S(l,b,\lambda)$, consists of two components: the first is a
discrete component, which includes all bright sources that exceed the local 
background level by a chosen amount, depending on wavelength; and the second 
is a smooth component, consisting of unresolved stellar sources which can be
represented by using a statistical model for calculating the brightness of the
Galaxy. 

Rather than modeling and removing each of the bright sources at all
wavelengths, the locations of these sources are simply blanked and excluded
from any further processing or analysis. Sources excluded by this procedure
consist of bright stars at short wavelengths, and diffuse sources such as the
Magellanic Clouds and bright star-forming regions in the Galactic plane
(e.g. the $\rho$ Oph and Orion regions), which appear at all wavelengths.
The procedure for constructing the bright source removed maps is described in
detail in Section 4.1. 

The unresolved stellar emission component is represented by a ``Faint Source
Model,'' which closely follows the ``SKY'' model described by Wainscoat et
al. (1992), and further developed by Cohen (1993a, 1994a, 1995). The SKY model
of Galactic structure was originally developed to fit ground-based $K$-band and
$V$-band and {\it IRAS} 12 and 25 $\micron$ Galactic source counts. The fitting
of source counts of known types of objects, instead of total intensity 
measurements, ensures that the
model intensity represents that of the Galactic sources, thereby excluding
any possible unresolved extragalactic contribution. 

To guarantee that the modeled ISM contribution to the sky brightness includes
only Galactic emission, we will write this emission component,
$G_I(l,b,\lambda)$, as a product of a spatially invariant spectral component
$R(\lambda)$ times a wavelength-independent spatial template.
The spatial template, such as the Galactic H I or CO column density, is 
associated with the Galactic gas phase component in which the dust resides.
The advantage of such a procedure is that the line emission from 
most extragalactic sources 
is redshifted to velocities higher than those at which Galactic line
emission is observed. Thus, the H I and CO data can be used as templates of 
the local (Galactic) IR emission of the ISM, without any contributions from 
extragalactic sources. In principle, $R(\lambda)$ can be obtained directly 
as the slope of a linear least squares fit to the correlation
of $I_{ZG_S}(l,b,\lambda)$ [$\equiv I_{obs}(l,b,\lambda) - Z(l,b,\lambda) -
G_S(l,b,\lambda)$] with the H I or CO spatial template for all wavelengths
$\lambda$. The intercept will then yield the value of $I_0(l,b,\lambda)$.
However, the main limitations to such a direct approach are: that there can be
relatively large spatial variations in $R(\lambda)$, 
the IR emission per H atom; and 
that high-quality H I and CO data sets have restricted sky coverage or 
different spatial resolution compared to the DIRBE data. These limitations can
be circumvented by using the 100 $\micron$ ISM emission $G_I(l,b,100$
$\micron)$ as the spatial template for ISM 
emission at the other wavelengths. The subtraction of the interstellar
component was therefore performed in the following manner. 
First, $R(\lambda)$ and $I_0(l,b,100$ $\micron)$ were
obtained as the slope and intercept of the correlation
between $I_{ZG_S}(l,b,100$ $\micron)$ and the H I spatial template.
This step allows us to separate the ISM emission, 
$G_I(l,b,100$ $\micron) \equiv I_{obs}(l,b,100$ $\micron) - 
Z(l,b,100$ $\micron) - I_0(l,b,100$ $\micron)$, from 
the potential CIB, $I_0(l,b,100$ $\micron)$. 
Then, we chose $G_I(l,b,100$ $\micron)$ as the spatial
template, and calculated $R(\lambda)$ by correlating this template with
$I_{ZG_S}(l,b,\lambda\neq100$ $\micron)$, so that $G_I(l,b,\lambda) =
R(\lambda) \times G_I(l,b,100$ $\micron)$. 

For the near-IR wavelengths ($\lambda \leq$ 4.9 $\micron$), the subtraction of
the starlight is not sufficiently accurate to yield values of $R(\lambda)$
and $I_0(l,b,\lambda)$ from direct correlations of the near-IR data with 
$G_I(l,b,100$ $\micron)$. However, the dust emission in the near-IR can be 
more easily distinguished from the stellar emission in
color-color plots (e.g., Arendt et al. 1994), as points that are displaced 
from the reddening line for stellar emission. This aspect of the subtraction
of the ISM-related near-IR emission is described in detail in Section 5.5. 

\section{DIRBE Data and the Infrared Emission from the Interplanetary Dust
Cloud}

The DIRBE instrument provides absolutely calibrated intensity sky maps in 10
broad bands at 1.25, 2.2, 3.5, and 4.9 $\micron$ (near-IR), 12, 25, 60, and 100
$\micron$ (mid-IR), and 140 and 240 $\micron$ (far-IR). The instrumental zero
point is established by chopping between the sky and a cold zero-flux
reference source at 32 Hz. Frequent observations of internal reference
sources and stable celestial sources allow corrections for small instrumental
gain instabilities on both short and long time scales. Measurements of 
a few well-calibrated celestial sources (Sirius for 1.25 -- 12 $\micron$,
NGC 7027 for 25 $\micron$, Uranus for 60 and 100 $\micron$, and Jupiter for 140
and 240 $\micron$) provide the absolute calibration of the DIRBE photometric
system. Further details on the calibration of the data can be found in the
{\it COBE} DIRBE Explanatory Supplement (1997). 

The DIRBE time-ordered data are pixelized and mapped in the {\it COBE} sky-cube
format ({\it COBE} DIRBE Explanatory Supplement 1997). All analysis is
performed on maps in this coordinate system. For illustrational purposes, maps
shown in this paper are reprojected into a Galactic Mollweide projection. The
Mollweide projection is an equal-area projection with the convenient
properties that longitudes are equally spaced (for a fixed latitude) 
and all lines of constant
latitude are straight horizontal lines (though unequally spaced). The DIRBE
surface brightness maps are presented in units of MJy/sr, which can be
converted to $\nu I_{\nu}$ intensities in units of 
nW m$^{-2}$ sr$^{-1}$ through multiplication by 3000/$\lambda$, 
where $\lambda$ is the wavelength of the band in $\micron$. 

The sky was scanned by DIRBE in a manner that provided highly redundant
coverage over a $60\arcdeg$ wide swath in a single week. The viewing swath
samples solar elongation angles from roughly $64\arcdeg$ to $124\arcdeg$, and
precesses with an annual period. The DIRBE beam is $\sim 0\fdg7 \times 0\fdg7$
at all wavelengths. A more complete description of the DIRBE instrument has
been given by Silverberg et al. (1993) and the {\it COBE} DIRBE Explanatory
Supplement (1997). The {\it COBE} mission is described by Boggess et al.
(1992). 

The highly redundant coverage per line of sight and wide range of solar 
elongation angles sampled provide an excellent database for purposes of 
modeling the IPD foreground in order to remove it from DIRBE 
observations. We use the parametric geometrical model described by Paper II
to compute the signal from the interplanetary dust cloud, 
denoted here as $Z$, for each line-of-sight (pixel) in the DIRBE maps. 
Components of the model are a smooth dust cloud, three pairs of dust 
bands, and a circumsolar ring in earth-orbit resonance. 
Details on the construction of this model and on the accuracy of its 
results are presented in Paper II.

Maps of the intensity after subtraction of the IPD emission and scattered 
sunlight are shown in Galactic Mollweide projection in 
Figure~\ref{zlsub_maps}. 
These data available as the `Zodi-Subtracted Mission Average (ZSMA)' maps, 
through the NSSDC {\it COBE} homepage website at 
${\tt http://www.gsfc.nasa.gov/astro/cobe/cobe\_home.html}$. 

\section{Infrared Emission from the Stellar Galactic Component}
 
\subsection{Bright Source Removal}

In order to model and remove the Galactic foreground emission on the largest
angular scales, we excluded small localized regions that are much brighter (and
often had distinctly different properties) than the average Galactic
foreground. These regions mainly consist of bright 
stars at the near-IR wavelengths, star forming regions at mid- and far-IR
wavelengths, and a few nearby external galaxies [See Odenwald, Newmark, \& 
Smoot (1997) for an analysis of the external galaxies visible to DIRBE]. The
locations of these bright sources have been individually blanked from the
maps, and were excluded from further analysis. Even though the 
intensities in the unblanked pixels are left unchanged, the mean intensity over large 
areas is reduced by the bright source blanking because the brightest pixels 
are now excluded.

A number of bright source
identification algorithms were tested, both for speed of execution and for
source detection accuracy. We adopted an algorithm based upon pixel brightness
relative to a locally determined smooth background. The smooth background
level for each pixel in the map was computed using a morphological filter (the
``opening'' operation; see Haralick, Sternberg \& Zhuang 1987), followed by a
linear smoothing. Pixels in the original map where the brightness exceeded the
background level by more than a fixed threshold were blanked. These
thresholds are: 15, 15, 15, 15, 85, 110 Jy for point sources at 1.25, 2.2,
3.5, 4.9, 12, and 25 $\micron$ respectively and 4.5 MJy sr$^{-1}$ = 135 nW
m$^{-2}$ sr$^{-1}$ for the more extended sources found 
at 100 $\micron$. The 60, 140, and 240 $\micron$ maps were
blanked at the same locations as the 100 $\micron$ map. Because the DIRBE beam
is larger than a standard pixel ($\sim0\fdg32\times0\fdg32$), the four nearest
neighbor pixels were blanked as well. Circular regions around the Large and
Small Magellanic Clouds were blanked at all wavelengths. The fraction of the
sky above $|b|=30\arcdeg$ that is excluded by the bright source removal ranges
from $\sim$35\% at the shortest wavelengths to $<$1\% at $\lambda$ $\geq$ 60
$\micron$. Figure~\ref{bs_map} shows the map of 2.2 $\micron$ intensity after
the contribution of the IPD has been removed and the bright source blanking
has been applied (cf. Fig.~\ref{zlsub_maps}). All subsequent intensity maps
will contain black regions indicating the locations where bright sources have
been blanked. 

This source removal technique does not work well at low Galactic latitudes,
where both source confusion and the curvature of the background are high. In
tests on simulated maps, however, the method was completely successful in
blanking sources above $|b|=20\arcdeg$. Further tests showed that
varying the threshold level by $\pm$ 20\% (a generous estimate of the map
noise) changed the median brightness of the sky above $|b|=15\arcdeg$
by well under one percent at 1.25 and 2.2 $\micron$, and even less at the
other wavelengths. 

\subsection{The Faint Source Model}

After the bright sources are removed by blanking, most of the residual diffuse
near-IR emission is starlight from unresolved sources. These sources are
sufficiently numerous and evenly distributed that a statistical model well
represents their collective emission. To remove the Galactic diffuse stellar
emission component we constructed a ``Faint Source Model'' (FSM) of the
integrated brightnesses in the DIRBE 1.25, 2.2, 3.5, 4.9, 12 and 25 
$\micron$ bands, based on the statistical model developed by Wainscoat et
al. (1992). Subsequent improvements to the Wainscoat et al. ``SKY'' model by
Cohen (1993a, 1994a, 1995) are generally not included in the FSM because they
do not pertain to emission at near-IR wavelengths or because they are not 
specified in detail such that they can be accurately reproduced within the
framework of the FSM. The FSM can be used to calculate the surface brightness
in any direction of the sky by integrating the flux emitted by the 
various stellar components along that line of sight. If the Galaxy consisted
of a single type of star, with a corresponding 
absolute magnitude $M_{\nu}$, then
the sky brightness along a line of sight would be given by: 
\begin{equation}
I_{\nu} = \frac{L_{0\nu} 10^{-0.4M_{\nu}}}{4\pi} \int^{\infty}_0 
e^{-\tau(s)} n(x,y,z) ds
\label{eq_fsm}
\end{equation}
where $n(x, y, z)$ is the number density of sources at a given position in
the Galaxy which is integrated along the line of sight $s$, $\tau(s)$ is the
optical depth along the line of sight, and $L_{0\nu}$ is 
the luminosity of a zero magnitude star. 
A more realistic representation of the sky brightness
must include a summation over many different
stellar types and their spatial distributions. The simple expression
(Eq.~\ref{eq_fsm}) is then replaced by the more complicated one: 
\begin{equation}
I_{\nu} = \frac{1}{4\pi} \sum_i \sum_j \int^{\infty}_0 e^{-\tau(s)} 
n(x,y,z,i,j) L_{0\nu}(i) \int_{\Delta M_{\nu}(i)} \omega(M_{\nu},i) 
10^{-0.4M_{\nu}(i)} dM_{\nu} ds
\end{equation}
where the summations over $i$ and $j$ represent, respectively, those over
stellar types, and over Galactic structural components, and the integral over
$M_{\nu}$ represents the fact that stars of a given type have an intrinsic
dispersion $\Delta M_{\nu}(i)$ in absolute 
magnitudes centered on $M_{\nu}(i)$ with a probability $\omega(M_{\nu},i)$.
The FSM follows the basic form of the Wainscoat et al. and SKY
models, including 5 structural components for the Galaxy (disk, spiral arms,
molecular ring, bulge, and halo), 87 source types, each with a dispersion of
absolute magnitudes, and interstellar extinction from dust in an exponential
(in radius and scale height, $z$) disk. The spiral arm component includes 
the local spur as defined by Cohen (1994a).

We have made several modifications to our FSM, deviating from the SKY model,
either out of necessity or to produce results more suited for comparison with
the DIRBE data: 

a) Since source count models cannot accurately represent the brightest point
sources, which are unevenly distributed across the sky, the FSM was only
integrated over stars fainter than those that were previously 
blanked from the DIRBE maps. The value of the brightness limit at which the
bright source blanking stops and the Faint Source Model begins is not a
significant source of uncertainty. Changes in the limits by 20\% produce
$<$5\% changes in the Faint Source Model intensities, which are largely offset
by complementary changes in the flux excluded by the bright source blanking
(Section 4.1.). 

b) We have increased the spatial resolution of the model and calculated
the sky brightness at the resolution of the DIRBE maps. 

c) The halo is described by Wainscoat et al. (1992) only as an $R^{1/4}$ law in
projected surface brightness. Therefore we have adapted the formulation for
the volume density presented by Young (1976) for use in our Faint Source
Model. At high Galactic latitudes, the halo is the third most important
component of the Faint Source Model after the disk and the spiral arms. 

d) We omitted the extragalactic component of the SKY model, which contributes
mainly at faint magnitudes at 25 $\micron$ (Cohen 1994a). 

e) The position of the Sun was set at 18 pc above the midplane of the Galactic
disk. This was determined during preliminary trials, by requiring equal 
brightnesses at the north and south Galactic poles after subtraction of the 
FSM in the near IR bands. This value is independently supported by analysis
done with the SKY model (Cohen 1995).

f) For 3.5 and 4.9 $\micron$ absolute magnitudes, which are not represented in
the Wainscoat et al. (1992) source table, we obtained approximate magnitudes
for the various source types by extrapolating from the $J$ magnitudes using
$V-J, V-L,$ and $V-M$ stellar colors from Johnson (1966). 

g) The conversion from magnitudes to fluxes was done in a manner consistent
with the absolute calibration of the DIRBE data (see {\it COBE} DIRBE
Explanatory Supplement, 1997). The intensities of the FSM at 2.2 $\micron$
were multiplied by a factor of 0.963 to 
account for differences between the color
corrections of Sirius (which was used as the DIRBE 
absolute calibrator) and those of
K and M giants which dominate the emission of the FSM and have a CO absorption
band partially within the 2.2 $\micron$ DIRBE passband. There exist
uncertainties of 10 -- 15\% in the model arising from details of the absolute
calibration of the model and the differences and treatment of the various
broadband filter responses (Cohen 1993b). 

To check our FSM, we compared the results of our calculations
to the average surface brightnesses of 238 large zones covering the entire sky
as calculated by the SKY model (Cohen 1994b; see Table 7 of
Wainscoat et al. (1992) for the zone boundaries). Our Faint
Source Model reproduces the mean intensities calculated with the SKY model to
within $\sim$5\% for most of the zones. The zones that are not well matched are
all low-latitude zones below $|b|=20\arcdeg$ (except for the zone
containing the LMC and another near the Taurus region) which have been
adjusted in the SKY model, but not in the FSM, for various localized features
within the Galaxy (Cohen 1994a). 

We have also compared the star counts expected
from our FSM with the star counts from the prototype 2MASS survey in 7 fields
at $J$ and $K_S$ (Skrutskie 1996). The fields are all in
the first Galactic quadrant and span the range $8\arcdeg < b < 87\arcdeg$. 
The FSM star counts resemble the 2MASS star counts, although about
half the fields show statistically significant differences, mostly 
at magnitudes $J > 12$ and $K_S > 12$. However, these fainter
stars contribute a small enough fraction of total emission from these fields
that the integrated brightnesses of the stars in the 2MASS fields are
consistent with the predictions of the Faint Source Model. This comparison is
limited by the small size of the fields (0.12 -- 7 deg$^2$) 
and thus the large statistical uncertainties on the star counts. 

Figure~\ref{fsm_map} shows the 2.2 $\micron$ sky brightness calculated with
the FSM. The model predicts the mean intensity in each pixel, and
thus varies smoothly over the sky. The actual sky is not expected to be as
smooth as the model because random Poisson deviations from the mean number of
stars/magnitude in each pixel create intensity fluctuations. Faint
discontinuities in the model are caused by the artificially sharp edges of the
spiral arms, as defined in the model (Wainscoat et al. 1992; Cohen 1994a). 

Maps of the FSM intensities at 1.25 -- 25 $\micron$ can be obtained 
from the NSSDC {\it COBE} homepage website at 
${\tt http://www.gsfc.nasa.gov/astro/cobe/cobe\_home.html}$. 

\subsection{The Residual Emission}

Figure~\ref{gfs_map} shows the 1.25 -- 4.9 $\micron$ residual emission 
($I_{ZG_S} \equiv I_{obs} - Z - G_S$) after the removal of 
stars and other compact objects with the bright source blanking and 
subtraction of the FSM. At these wavelengths, the subtraction
of the stellar emission is very effective at removing the brightness gradients
at high Galactic latitudes. The maps are mottled by the emission of the
remaining faint, marginally resolved stars, which lie below the bright source
threshold. The average emission of these remaining stars is included in 
the smooth Faint Source Model. Therefore, over sufficiently large regions this
stellar confusion acts as an additional noise term in the 
residual maps. At wavelengths $\leq$ 3.5 $\micron$ and low
Galactic latitudes ($|b|<10\arcdeg$), the FSM tends to overestimate the actual
sky brightness, leading to negative residuals. A longitudinally anti-symmetric
residual is present at the location of the Galactic bulge (most clearly seen
in the 4.9 $\micron$ map of 
Fig.~\ref{gfs_map}), which suggests that improvements could result if a
bar-like model were used for the bulge (e.g., Dwek et al. 1995; Blitz \&
Spergel 1991). Some relatively large differences between the model and the
data are found at locations where our implementation of the Faint Source Model
has omitted particular adjustments for specific disk and spiral arm features
(Cohen 1994a). Residual defects from the subtraction of the IPD emission are
seen clearly at wavelengths $\geq$ 3.5 $\micron$ as S-shaped bands along the
ecliptic. 

In the mid-IR (after removal of IPD emission), the faint source emission
contributes $<$40\% of the observed brightness toward the inner Galaxy at low
latitudes, and less than 5\% and 1\% for $|b|>30\arcdeg$ at 12 and 25 
$\micron$, respectively. The subtraction of the Faint Source Model in these
bands is only apparent in the inner Galaxy, and residual intensity maps look
nearly the same as those in Figure~\ref{zlsub_maps}. The residual emission at 
wavelengths $\geq$ 12 $\micron$ is clearly dominated by the presence of ISM
emission, $G_I(l,b,\lambda)$. 

The FSM was not optimized to fit the DIRBE data, but is calculated from a
fixed prescription based on a model designed to fit observed source counts in
selected directions in the sky. It is therefore important to examine to what
extent the removal of Galactic starlight with the FSM has under- or
over-subtracted stellar emission from the maps. One test designed to look for
such effects is to search for structure in the residual map that may be
correlated with that of the FSM, or Galactic latitude. 

Figures~\ref{gfs_b_grad} and~\ref{gfs_beta_grad} illustrate the gradients of
$I_{ZG_S}(l,b,\lambda)$ as a function of Galactic and ecliptic latitude,
respectively, for the 1.25, 2.2, 3.5, and 4.9 $\micron$ data.
Table~\ref{tab_fsm_grad} lists the gradient and correlation coefficient for
$I_{ZG_S}(l,b,\lambda)$ as a function of $\csc(|b|)$. The figures and table
both show that the Faint Source model removes much of the Galactic gradient in
the near-IR emission. Subtraction of the model also reduces the longitudinal
variation of the near-IR sky brightness, as evidenced by the reduced
dispersions at lower latitudes in Figures~\ref{gfs_b_grad} 
and~\ref{gfs_beta_grad}. At 4.9 $\micron$, the residual IPD gradient visible 
in Figure \ref{gfs_beta_grad} is responsible for the structure in the residual 
emission as a function of $\csc(|b|)$ that is not observed at shorter 
wavelengths (Fig. \ref{gfs_b_grad}).

\section{Infrared Emission from the Galactic Interstellar Medium}

The last foreground component we need to remove is the emission from the
Galactic ISM. As outlined in Section 2, we use the 100 $\micron$ DIRBE data
(after removal of the IPD emission) to derive a spatial template of the ISM.
We correlate the 100 $\micron$ data with tracers of the gas phase of the ISM,
and subtract emission that is uncorrelated with the ISM to form the 100
$\micron$ ISM template. The external data 
used to trace the gas phase of the ISM are introduced in Section 5.1. Then
we derive the 100 $\micron$ emission per H I column density, and 
assess the possibilities that some of the emission comes from dust within the
ionized and molecular phases of the ISM (\S5.2). 
After subtracting the 100 $\micron$
emission that is not associated with the ISM, we correlate the 100 $\micron$
ISM template against the emission at other infrared wavelengths to derive the
ISM intensity at those wavelengths (Section 5.3). The remaining
subsections describe modified procedures that we apply to identify the
ISM emission at the longest (240 $\micron$) and shortest (3.5 and 4.9
$\micron$) wavelengths. 

The 100 $\micron$ map is chosen as the template of the high latitude ISM
because: (a) it provides full sky coverage; 
(b) it has better sensitivity than the maps at longer wavelengths
where noisier bolometric detectors were used; (c) it is less affected by
errors in the removal of IPD emission than the shorter wavelength maps; and
(d) it represents a compromise between potential differences in spatial
structure of the shorter and longer wavelength ISM emission. One drawback to 
this choice is that larger and a greater number of corrections for instrument
temperature, charged particle radiation, and photon induced responsivity
changes were required at 100 $\micron$ than at any other wavelength (see {\it
COBE} DIRBE Explanatory Supplement 1997). 

\subsection{External Data Sets}

In order to identify that portion of the 100 $\micron$ intensity (remaining
after subtraction of IPD emission) that arises from the ISM, we need to
compare the DIRBE data with other data sets that can be safely assumed to
trace only Galactic ISM emission. 

\subsubsection{H I Data}

Three H I 21 cm line surveys have been used for comparison with the long
wavelength DIRBE data. The first is the AT\&T Bell Laboratories survey of the
sky north of $\delta = -40\arcdeg$ at an angular resolution of about $2\arcdeg$
(Stark et al. 1992). The velocity-integrated ($|v|\leq327$ km s$^{-1}$) line
intensities were converted to H I column densities assuming the line emission
is optically thin, and a map of column density was produced in {\it COBE}
sky-cube
format ({\it COBE} DIRBE Explanatory Supplement 1997). The Bell Laboratories
20 foot horn reflector has very low far-sidelobe response. Lockman, Jahoda,
and McCammon (1986) estimated that the maximum far-sidelobe contribution to
observed column density is $8\times10^{18}$ cm$^{-2}$, and the typical
contribution is $0.5\times10^{18}$ cm$^{-2}$. The second data set consists of
observations of an $8\arcdeg \times 9\arcdeg$ region centered on the North
Ecliptic Pole ($l=96\arcdeg, b=+30\arcdeg$) at $21\arcmin$ resolution (Elvis,
Lockman, and Fassnacht 1994). Elvis et al. used the Bell Laboratories survey
to correct their spectra for stray radiation, and estimated that the 1$\sigma$
random error in column density is $1\times10^{19}$ cm$^{-2}$. The third data
set covers a 300 deg$^2$ region in Ursa Major surrounding the direction of
lowest H I column density ($l=150\arcdeg, b=+53\arcdeg$) at a resolution of
$21\arcmin$ or better (Snowden et al. 1994). Snowden et al. also corrected
their spectra for stray radiation, and estimated 1$\sigma$ errors ranging from
$0.5\times10^{19}$ cm$^{-2}$ to $1.0\times10^{19}$ cm$^{-2}$.
Velocity-integrated maps of the North Ecliptic Pole and Ursa Major regions
were obtained from F. J. Lockman (private communication). The velocity
ranges of $-150$ to 150 km s$^{-1}$ for the North Ecliptic Pole and
$-150$ to 100 km s$^{-1}$ for Ursa Major include all significant Galactic
emission. The maps were
degraded to DIRBE resolution and reprojected to {\it COBE} sky-cube format. 

\subsubsection{CO Data}

The CO data used in this analysis comprise the $^{12}$CO survey of the
Ophiuchus region by de Geus et al. (1990), and the $^{12}$CO surveys of the
Polaris flare, Ursa Major, and Camelopardalis regions by Heithausen et al.
(1993). These surveys, which cover a total area of 900 deg$^2$ at high 
($|b|>15\arcdeg$) Galactic latitude, were taken with the 1.2~m
millimeter-wave telescope initially located at Columbia University in New
York City, and later at the Harvard-Smithsonian Center for Astrophysics in
Cambridge, MA, or the nearly-identical Columbia Southern Millimeter-Wave
Telescope in Cerro Tololo, Chile. Both telescopes have an angular resolution 
of 8\farcm7 at the $^{12}$CO J=$1\rightarrow0$ line frequency observed. The
surveys were used to produce a velocity-integrated $^{12}$CO map degraded to
the resolution of the Bell Labs H I survey. 

\subsubsection{H II Data}

The analysis presented in the Appendix makes use of pulsar dispersion measures
from Taylor, Manchester, and Lyne (1993) and Camilo and Nice (1995), and
H$\alpha$ intensities observed by Reynolds (1980, 1984, 1985, 1991b) using a 
Fabry-Perot spectrometer with a beam $50\arcmin$ in diameter. 

\subsection{Correlation of 100 $\micron$ Intensity with Gas Phases of the ISM}

Our primary identification of the 100 $\micron$ ISM emission is made through
correlation with H I column densities toward the North Ecliptic Pole and the
region of minimum H I column density (the Lockman Hole). The choice of these
locations is determined by the low column densities, the availability of high
quality H I data sets, and the ability to estimate or limit contributions from
the molecular and ionized components of the ISM. The slope of the correlation
is the average 
100 $\micron$ emissivity per H I column density. The intercept of the
correlation is the mean intensity that is {\it not} 
associated with the H I component of the ISM. We also estimate the intensity 
of 100 $\micron$ emission that may be associated with the ionized and 
molecular phases of the ISM. However, direct correlations between the 100
$\micron$ emission and combinations of $N$(H I) and $N$(H II) or $N$(H$_{2}$)
are found to be relatively ineffective because of low column density and/or
emissivity of the molecular and ionized components. 

\subsubsection{Emission from Dust in H I}
 
To determine the association of the 100 $\micron$ emission with the neutral 
atomic ISM, we correlate the H I column density, $N$(H I), against the 100
$\micron$ intensity after subtraction of IPD emission (and, in principle,
any stellar emission), $I_{ZG_S}(100~\micron)$, 
\begin{equation}
\nu I_{ZG_S}(100~\micron) = A~N({\rm H~I}) + \nu I_0(100~\micron)
\end{equation}
to solve for the emissivity of the neutral ISM, $A$, and the mean intensity of 
emission that is uncorrelated with H I, $\nu I_0(100~\micron)$.
Figure~\ref{b10_hi}a shows the pixel-by-pixel comparison for the
$8\arcdeg\times 9\arcdeg$ region centered on the North Ecliptic Pole (NEP).
Figure~\ref{b10_hi}b shows the comparison for the 300 deg$^2$ region
surrounding the region of lowest H I column density (the Lockman Hole).
Figure~\ref{b10_hi}c shows the comparison for a region at $\vert{\it b}\vert$
$>$ 25\arcdeg, $\vert{\it \beta}\vert$ $>$ 25\arcdeg, and declination $\delta >
-40\arcdeg$, using 100 $\micron$ data smoothed to the resolution of the Bell
Laboratories survey. This very large region is included to provide an 
indication of the extent to which the Lockman Hole and NEP regions are 
typical of other high latitude regions. 
Data in a $3\times3$ DIRBE pixel patch centered on the
source RAFGL 5429 were excluded from Figure~\ref{b10_hi}a, and data in
$3\times3$ pixel patches around five galaxies (NGC 3079, 3310, 3556, 3690, and
4102) were excluded from Figure~\ref{b10_hi}b. The solid lines show linear
fits to the data that minimize $\chi^2$ calculated using measurement errors
in both variables (Press et al. 1992). Each plot shows excess 100 $\micron$
emission at the highest H I column densities. These data were excluded from
the fitting by rejecting all data 
above a cut line that is perpendicular to the fit line on a
plot where each variable is divided by its mean measurement error. The
adopted cut lines are shown as short diagonal lines 
in Figures~\ref{b10_hi}a--c. The
cut line intersects the fit line at $N$(H~I)=$5.0\times10^{20}$ cm$^{-2}$ for
the NEP region, at $1.5\times10^{20}$ cm$^{-2}$ for the Lockman Hole region,
and at $3.0\times10^{20}$ cm$^{-2}$ for the region of Figure~\ref{b10_hi}c.
Each fit was done iteratively until the fit parameters and the data excluded
by the cut stopped changing.
The derived intercept values are not sensitive to the exact location of
the cut lines. For the Lockman Hole region, for example, the derived
intercept varied by less than 0.7 nW m$^{-2}$ sr$^{-1}$ 
for a series of test fits
in which the location of the cut line intersection ranged
from $1.0\times10^{20}$ to $2.0\times10^{20}$ cm$^{-2}$.
Figure~\ref{b10_hi_binned} shows a comparison of
the fit line with data averaged within uniformly spaced bins along the fit
line for each of the three regions. In each region, the bin-averaged data are
consistent with a linear relation over the range of column density used for
the fitting. 
These correlations begin to deviate from linearity 
at higher column densities because of the contribution of emission from dust 
associated with molecular H$_2$ (e.g. Reach, Koo, \& Heiles 1994; Reach, 
Wall, \& Odegard 1998). 
 
The fit parameters (slopes, $A$, and intercepts, $\nu I_0(100$ $\micron)$) 
are listed in Table~\ref{tab_b8_hi}. The tabulated uncertainties are 
formal errors determined from the 68\% joint confidence region in parameter
space. The average of the intercepts for the NEP and Lockman Hole regions,
19.8 nW m$^{-2}$ sr$^{-1}$, was subtracted from the 100 $\micron$ 
IPD-subtracted skymap to form the spatial template of 100 $\micron$ ISM emission
(see Section 5.3). 
The difference between the intercepts for the NEP and Lockman Hole regions is
5.0 nW m$^{-2}$ sr$^{-1}$. This value was adopted as the 
systematic uncertainty for the zero level of the template. 

Sources of systematic error that may affect the zero level of the 100 
$\micron$ ISM template 
include errors in subtraction of the IPD emission and errors due to 
variations in 100 $\micron$ emissivity per neutral H atom
(the ${\nu}I_{ZG_S}$(100 $\micron$)/$N$(H I)
ratio for emission originating from the Galaxy) within the NEP or Lockman 
Hole regions. The latter could be caused by variations in dust temperature,
grain composition, grain size distribution, or dust-to-gas mass ratio within 
the neutral atomic gas phase, or by emission from dust associated with 
ionized or molecular gas. If the 100 $\micron$
emissivity per neutral H atom increases with increasing H I column density 
within a region, the intercept of the linear ${\nu}I_{ZG_S}$(100 $\micron$) $-$ 
$N$(H I) fit would underestimate the zero level of the emission from the ISM, 
and if it decreases with H I column density the intercept would overestimate
the zero level. However, such systematic variations would tend to produce
nonlinear ${\nu}I_{ZG_S}$(100 $\micron$) $-$ $N$(H I) relations. A quadratic
dependence might be expected for a cloud containing H$_{2}$ (e.g., Dall'Oglio,
et al. 1985; Reach, Koo, \& Heiles 1994), but specific functional forms are
not predicted for other causes of variation in emissivity per neutral H atom.
A simple interpretation of the linear relations found for the NEP and
Lockman Hole is that systematic variations of emissivity per neutral H atom
with H I column density are not important at low H I column densities in these 
regions.

The difference between the intercepts found for the NEP and Lockman Hole
regions can probably be attributed to error in subtraction of IPD emission.
Evidence for this can be seen in Figure~\ref{lat_trends}, which shows that a 
weak ecliptic latitude dependence remains
in the 100 $\micron$ data after subtraction of the IPD emission model. The
figure shows a similar ecliptic latitude dependence for data within four
narrow ranges of H I column density. The intensity difference between
$\beta=45\arcdeg$ and $75\arcdeg$ in Figure~\ref{lat_trends}
is comparable to the difference in intercepts
derived for the Lockman Hole ($\beta = 45\arcdeg$) and NEP 
($\beta = 90\arcdeg$) . 

\subsubsection{Possible Emission from Dust in H II}

Because the preceding correlations only identify the 100 $\micron$ emission
that is correlated with the neutral H I phase of the ISM, we need to estimate 
any error in the inferred zero
level of the 100 $\micron$ ISM template that may be caused by dust emission
associated with the ionized phase of the ISM. The exact distributions of
ionized hydrogen column density within the NEP and Lockman Hole regions are
not known. However, for the Lockman Hole, available data can be used with some
assumptions to estimate the possible error of the zero level.
We initially assume that the 100 $\micron$ emissivity per H nucleus within
the H II gas phase equals the slope of the ${\nu}I_{ZG_S}$(100 $\micron$) $-$
$N$(H I) relation, and that the H II column density is constant within the
region. Under these assumptions, the ${\nu}I_{ZG_S}$(100 $\micron$) $-$ $N$(H
I) relation would still be linear but its intercept ($I_0$) would 
overestimate the zero level of emission from the ISM by $N$(H II) times the
emissivity per H nucleus. 

Table~\ref{tab_hi_h2} lists information on ionized H column density for the
Lockman Hole based on a pulsar dispersion measure and two H$\alpha$
observations. We are not aware of any useful data on ionized gas in the NEP
region. The dispersion measure $DM \equiv \int n_{e} ds$
 of PSR J1012+5307 provides a lower limit to
the total $N$(H II) along the line of sight. Estimates of the distance of the
pulsar from the midplane are $z$=460 pc from photometry of its companion,
assuming it is a $0.15 M_{\sun}$ white dwarf (Lorimer et al. 1995; Halpern
1996), and $z$=400 pc from the dispersion measure and the Galactic electron
density model of Taylor \& Cordes (1993). Estimates of the exponential scale
height of the ionized medium in the vicinity of the Sun range from 670 to 910
pc (Reynolds 1991a; Nordgren, Cordes, \& Terzian 1992; Taylor \& Cordes
1993). If the ionized gas in the Lockman Hole region has an exponential $z$
distribution with a scale height in this range, and the pulsar is at $z$=460 pc,
the total $N$(H II) in the direction of the pulsar would be 2 to 2.5 times the
dispersion measure, or about $0.6 \times 10^{20}$ cm$^{-2}$. 

The H$\alpha$ observations provide a measure of $\int T^{-0.92} n_{e}^{2} ds$
(e.g., Reynolds 1992), since extinction is negligible for lines of sight with
such low H I column densities.
H II column densities have been estimated from the H$\alpha$ data in 
Table~\ref{tab_hi_h2} 
using a conversion factor of $I$(H$\alpha$)/$N$(H II)=0.75 Rayleighs/$10^{20}$
cm$^{-2}$ (1 Rayleigh = 10$^{6}$/4$\pi$ photons s$^{-1}$ cm$^{-2}$ sr$^{-1}$ =
0.24 nW m$^{-2}$ sr$^{-1}$ at H$\alpha$).
This value was adopted based on the comparison of H$\alpha$ and
dispersion measure data shown in Figure~\ref{ha_dm} for high $\vert z \vert$
pulsars. The figure shows that $I$(H$\alpha)/DM$ may increase with increasing H
I column density. The $I$(H$\alpha)/DM$ ratio at the lowest H I column densities
in the figure was adopted as the conversion factor 
for the Lockman Hole data. The adopted value corresponds to a characteristic
electron density, $n_c \equiv \int n_e^2 ds/\int n_e ds$, of 0.05 cm$^{-3}$ for
an electron temperature of 8000 K. For comparison, Reynolds (1991a) found a 
mean $n_c$ of 0.08 cm$^{-3}$ for lines of sight toward pulsars in 4 globular 
clusters at $\vert z \vert >$ 4 kpc.

The $N$(H II) values estimated from the H$\alpha$ observations are comparable 
to or smaller than the pulsar's dispersion measure. Even though we consider 
the $N$(H II) estimates from the H$\alpha$ data to be more reliable than that 
derived from the dispersion measure, pulsar $z$ distance, and
scale height of ionized gas, to be conservative we adopt an intermediate value
of $N$(H II)=0.4$\times10^{20}$ cm$^{-2}$ as representative of the Lockman Hole
region. Dust associated with ionized gas at this column density
would have a 100 $\micron$ intensity of 7 nW m$^{-2}$ sr$^{-1}$ under the
assumptions described above, and the ${\nu}I_{ZG_S}$(100 $\micron$) $-$ $N$(H I)
intercept $I_0$ would overestimate the zero level of the ISM emission by this
amount, about 1/3 of the total. 

However, this error estimate is probably too large, for two
reasons. (1) The analysis presented in the Appendix suggests that the diffuse
ionized gas at high galactic latitudes has a 100 $\micron$ emissivity per H 
nucleus that is smaller than that of the H I gas. A 3$\sigma$ upper limit of
3/4 times the emissivity per H nucleus of the H I gas was found for a 
$10\arcdeg\times12\arcdeg$ region at $l=144\arcdeg$, $b=-21\arcdeg$. 
Analysis by Fixsen et al. (1998) using $COBE$/FIRAS data also indicates that 
the emissivity of the ionized gas is small. They correlate far-IR emission 
against a combination of [C II] line emission, as a tracer of the ionized ISM, 
and linear and quadratic H I column density, as a tracer of the neutral ISM. 
At high latitudes, they find little or no far-IR continuum emission 
correlated with the [C II] line emission.
(2) It is likely that H II column density is not constant in the Lockman Hole
region, but is correlated to some degree with H I column density. Such 
correlations are found in the Appendix for data along lines of sight toward 
two samples of high $\vert z \vert$ pulsars. 
If such a correlation exists in the Lockman hole, the error estimate for
the 100 $\micron$ zero point should be calculated using only the component of
$N$(H II) that is not correlated with $N$(H I).
Further evidence for correlation between ionized and atomic gas has been
shown for the $10\arcdeg\times12\arcdeg$ region at $l=144\arcdeg$, 
$b=-21\arcdeg$
by Reynolds et al. (1995). They found that at least 30 percent of the total 
H$\alpha$ emission and 10 to 30 percent of the total 21 cm line emission are 
kinematically and spatially associated in clouds containing both neutral and 
ionized gas. For these reasons, we adopt 4 nW m$^{-2}$ sr$^{-1}$, about 1/5
of the total, as the maximum possible systematic error in the zero level
of the 100 $\micron$ ISM template that may result from H II associated dust.

\subsubsection{Possible Emission from Dust in H$_{2}$}

We additionally need to estimate the error in the zero level of the 100 
$\micron$ ISM template that may be caused by emission from dust associated 
with the molecular ISM.
Table~\ref{tab_hi_h2} lists 3-sigma upper limits on H$_{2}$ column
density for both the Lockman Hole and NEP regions, based on upper limits to 
$^{12}$CO J=1--0 
line emission within a $10\arcdeg\times10\arcdeg$ portion of the Lockman Hole
region and toward some H I emission peaks in the NEP region. [CO emission has
been detected toward the 100 $\micron$ brightness peaks of a few cirrus clouds
elsewhere in the Lockman Hole region by Heiles, Reach, \& Koo (1988), Stacy et
al. (1991), and Reach, Koo, \& Heiles (1994). The H I column densities at
these positions are $1.9\times 10^{20}$ cm$^{-2}$ or greater, so they were
excluded from the ${\nu}I_{ZG_S}$(100 $\micron$) vs. 
$N$(H I) fitting.] Following
Elvis et al. (1994), limits on observed line emission were converted to limits
on H$_{2}$ column density assuming a 1 km s$^{-1}$ line width, typical of
diffuse clouds, and a CO intensity to $N$(H$_{2}$) conversion ratio of $X_{CO} <
4\times10^{20}$ mol cm$^{-2}$ (K km s$^{-1}$)$^{-1}$. Most determinations of
$X_{CO}$ for translucent molecular clouds (clouds with visual extinction in
the range $1 < A_{V} < 5$ mag) are consistent with this upper limit (Magnani
\& Onello 1995, and references therein). Dust associated with molecular gas at
or below the $N$(H$_{2}$) limits in Table~\ref{tab_hi_h2} would produce
100 $\micron$ intensities of $<$42 and $<$23 nW m$^{-2}$ sr$^{-1}$ for the
Lockman Hole and NEP, respectively, assuming constant $N$(H$_{2}$) across these 
regions and emissivity per H nucleus equal to the slope of the
${\nu}I_{ZG_S}$(100 $\micron$)--$N$(H I) relation. 
These conclusions for the Lockman Hole and NEP can be extended to cover 
the entire north Galactic hemisphere, since a recent CO survey for 
$b\geq 30\arcdeg$ by Hartmann, Magnani, \& 
Thaddeus (1998) has comparable sensitivity to the measurements in Table 3
and finds that detectable CO has a filling factor of 0.004 -- 0.006.

For the Lockman Hole region, the H$_{2}$ column density and the possible 100
$\micron$ emission from dust associated with H$_{2}$ are probably much smaller
than allowed by the limit on CO line intensity. Ultraviolet absorption line
observations toward early type stars have shown that a transition in the
fraction of H in molecular form occurs at color excess $E(B-V)\simeq$0.08 mag,
or at $N$(H I) + 2$N$(H$_{2}$)$\simeq 5\times 10^{20}$ atoms cm$^{-2}$, above
which more than 1\% of H atoms are in the form of H$_{2}$ (Savage et al.
1977). Studies of the H I--H$_{2}$ phase transition based on H I, CO, and 100
$\micron$ observations of high latitude molecular clouds 
are consistent with this
(Reach, Koo, \& Heiles 1994; Gir, Blitz, \& Magnani 1994). The $N$(H$_{2}$)
limit in Table~\ref{tab_hi_h2}, together with the maximum H I column density
used in the ${\nu}I_{ZG_S}$(100 $\micron$)--$N$(H I) fitting, yields $N$(H I) +
2$N$(H$_{2}$)$< 4.0\times10^{20}$ atoms cm$^{-2}$, suggesting that the gas in
the Lockman Hole is below the threshold for significant fractional H$_{2}$
abundance. However, the $N$(H$_{2}$) limit may not be valid if H$_{2}$ is
present in clouds that are too diffuse for CO to exist. 

Estimates of visual extinction toward the Lockman Hole give another
indication that the H$_{2}$ column density is very low. The Galactic reddening
map of Burstein \& Heiles (1982), which is based on maps of galaxy counts and
H I column density and is calibrated using reddening measurements for globular
cluster stars and RR Lyrae stars, gives a mean $E(B-V)$ of $-0.006\pm0.01$ for
the Lockman Hole region. The reddening map of Holmberg (1974), based on counts
of distant Zwicky galaxy clusters and calibrated assuming $A_{B}$=0.25 at the
galactic poles, gives $E(B-V)=0.035\pm0.014$. The difference between these
results is primarily due to differences in calibration; if the Holmberg map
were recalibrated to the mean polar reddening of Burstein and Heiles, the
results would agree within the quoted relative errors. Recent determinations
of polar absorption by different methods are in the range $0< A_{B}< 0.21$ (de
Vaucouleurs 1995; Knude 1996, and references therein) so the mean $E(B-V)$ for
the Lockman Hole region is probably less than 0.035 and the reddening for most
or all of the positions used for the 
${\nu}I_{ZG_S}$(100 $\micron$)--$N$(H I) fitting
is probably below $E(B-V)=0.08$. Adopting the largest fractional H$_{2}$
abundance measured by Savage et al. (1977) for $E(B-V) < 0.08$ would give an 
average
H$_{2}$ column density of $5\times10^{15}$ mol cm$^{-2}$ for the Lockman Hole.
Assuming constant $N$(H$_{2}$) within the region and emissivity per H atom in
the H$_{2}$ phase equal to the slope 
of the ${\nu}I_{ZG_S}$(100 $\micron$) -- $N$(H I) relation, 
dust associated with molecular gas would give rise to a 100 
$\micron$ intensity of only 0.002 nW m$^{-2}$ sr$^{-1}$. 

As in the case for the ionized ISM, we have also attempted to determine the 
100 $\micron$ emissivity of the molecular ISM.
For each of the four regions covered by the above $^{12}$CO surveys (\S5.1.2),
we performed a least-squares fit of the equation 
\begin{equation}
\nu I_{ZG_S}(100~\micron) = A~N(H I) + B~W(CO) + \nu I_0(100~\micron)
\end{equation}
to the data by varying the parameters $A, B$, and $\nu I_0$, where $A$ is the
100 $\micron$ emissivity per H atom for the H I gas phase, $B$ is the product
of the 100 $\micron$ emissivity per molecule for the H$_2$ gas phase and the 
ratio of H$_2$ column density to $^{12}$CO intensity, and $\nu I_0$ represents any
background emission. In each case the resulting fit to the 100 $\micron$
fluxes is significantly poorer than those obtained in Figure~\ref{b10_hi},
where the contribution of dust associated with molecular gas to the 100
$\micron$ emission is believed to be negligible. The breakdown of the
infrared-gas column density correlation within high-latitude regions having
significant CO emission may result from one or more of the following: (1)
There may be a greater variability in the dust-to-gas mass ratio or dust
temperature within H$_2$ regions than within H I regions; (2) Regions with a
higher CO content may have more dust associated with ionized gas which is not
accounted for in the correlations. The photographic H$\alpha$ survey of Sivan
(1974) shows that a significant fraction of the Ophiuchus region contains
extended H II regions; (3) The dust column density within these regions may be
too low to effectively shield the CO molecule from dissociating UV radiation
(see, for example, Heithausen et al. 1993), and therefore the
velocity-integrated $^{12}$CO intensity may not accurately trace the H$_2$
column density along the line of sight. 

\subsection{Correlation of the Other DIRBE Bands with the 100 $\micron$
Emission} 

Having estimated the 100 $\micron$ emission that is not associated with the 
ISM ($\nu I_0 = 19.8$ nW m$^{-2}$ sr$^{-1}$), we construct the
$G_I(l,b,100$ $\micron)$ map from $I_{ZG_S}(l,b,100$ $\micron) - I_0(100$
$\micron)$, which is then used as the template for the ISM emission for
derivation of maps of the background emission at the other DIRBE wavelengths: 
\begin{equation}
I_0(l,b,\lambda) = I_{ZG_S}(l,b,\lambda) - R(\lambda) G_I(l,b,100~\micron).
\end{equation}

We fit the correlations of $I_{ZG_S}$ to $G_I$ to determine $R(\lambda)$, the 
$\lambda - 100$ $\micron$
color, from the slope of the correlation and a mean $I_0(l,b,\lambda)$ from the
intercept. These correlations and associated least-squares fits are shown in
Figure~\ref{bx_b8}. The slopes and their statistical
uncertainties are given in Table~\ref{tab_bx_b8}. For the far-IR bands, the
correlations are determined 
only for the regions where $|b|>45\arcdeg$. Using a cut at lower
latitudes resulted in a steeper slope, influenced by bright clouds found only
at the lower latitudes. A higher latitude cut resulted in a shallower slope,
possibly an artifact of the poorer correlation provided by the fainter, very
high latitude cirrus. We settled on the $45\arcdeg$ cut as a compromise
between the weaker correlations at higher latitudes, and the intrinsic color
temperature variations between the clouds visible at the lower latitudes. At
mid-IR wavelengths, residuals for the IPD removal (Paper II) were
too large to allow use of this entire region for correlation. Smaller regions
where the IPD and FSM removal have left no obvious structure were used: 
$|b|>30\arcdeg$ and $|\beta|>40\arcdeg$ at 60 $\micron$, and
$\beta > 70\arcdeg$ at 12 and 25 $\micron$. In the near-IR bands ($\lambda
\leq 4.9~\micron$), residuals from removal of starlight and IPD scattering and 
emission were too large to allow a direct correlation of these bands with the
100 $\micron$ ISM template (see Section 5.5). 

In using linear least-squares fits, it is assumed that the $R(\lambda)$ colors
of the ISM are constant at high latitudes. The validity of this assumption is
indicated by relatively large correlation coefficients, and by dispersions
about the least-squares fit relation which are no more than a factor of 2
greater than those expected from the uncertainties associated with the
IPD-subtracted data. These quantities are listed in Table~\ref{tab_bx_b8}. 
Another indication of the expected uniformity of the $R(\lambda)$ colors is
the observed Galactic temperature gradient reported by Sodroski et al.
(1994). Even though this radial temperature gradient leads to large color 
variations at low latitudes, at $|b|>45\arcdeg$ the
gradient implies large scale variation in the $R(240)$ color of only $\pm$4\% 
between the inner and outer Galaxy. The color variations at other wavelengths
should be even smaller, as the radial temperature gradients are smaller 
(Sodroski et al. 1987, 1989). In our present results any such 
large scale trends are
obscured by smaller scale cloud-to-cloud variations and therefore have no
direct impact on the derived background level. 

Scaling the 100 $\micron$ ISM template by the appropriate colors ($R(\lambda)$ 
from Table~\ref{tab_bx_b8})
and subtracting it from each of the mid- and far-IR wavelength maps yields the
maps shown in Figure~\ref{gcr_maps}. For the 100 $\micron$ residual map we
depict the subtraction of $1.84 \times 10^{-19}$ nW m$^{-2}$ sr$^{-1}$ cm$^2 
\times N$(H I) from the data (derived in
Section 5.2.1). These maps show the end result of our removal of both
IPD and Galactic emission. They should contain the full CIB signal as well as 
any residual emission left by imperfect foreground modeling.
Figures~\ref{gcr_b_grad} 
and~\ref{gcr_beta_grad} illustrate the gradients depicted in
the residual sky brightness of these maps as functions of Galactic and
ecliptic latitude. Table~\ref{tab_gcr_grad} lists these gradients and the
correlation coefficients with respect to $\csc(|b|)$. At 12 and 25 microns,
the subtraction of the ISM emission allows the residual gradients of
IPD emission to be seen and characterized more clearly, as indicated by 
Figures~\ref{gcr_maps}, \ref{gcr_b_grad},
and~\ref{gcr_beta_grad} and in Table~\ref{tab_gcr_grad}. At longer wavelengths,
the IPD residuals are weak compared to the residual ISM emission.

\subsection{A Two-Component Model for the 240 $\micron$ Emission}

The single component model of the ISM based on the 100
$\micron$ data removes much of the emission of the ISM at high latitudes.
However the residual maps (Fig.~\ref{gcr_maps}) show that at the far-IR
wavelengths a strong second component to the ISM emission is still present,
especially at lower latitudes ($|b|<30\arcdeg$). The 
similarity of the 140 and 240 $\micron$ residual maps suggests that a 
two-component model with templates formed from the 100 $\micron$ ISM emission
and the 140 $\micron$ residual could be effective at removing the 240
$\micron$ emission of the ISM. For a linear combination of these two
templates, we can substitute an equivalent linear combination of the 100
$\micron$ ISM emission and the 140 $\micron$ ISM emission. We use a linear
least squares fit to solve for the coefficients that scale the templates to
fit the 240 $\micron$ data. Performing this correlation over the entire
unblanked sky leads to a 240 $\micron$ ISM model of 
\begin{equation}
G_I(l,b,240~\micron) = -1.702 G_I(l,b,100~\micron) + 1.555 G_I(l,b,140~\micron),
\label{eq2color}
\end{equation}
where the ISM components at 100 and 140 $\micron$ have had constant terms 
(estimated from the one-component ISM subtraction residuals) of 
19.8 and 18.5 nW m$^{-2}$ sr$^{-1}$ removed respectively. This two-component 
model of the ISM emission will be referred to as the ``ISM2'' model to 
distinguish it from the one-component (ISM1) model described in \S 5.3.

A map of the residuals after subtraction of the ISM2 model from the
data is shown in Figure~\ref{gcr2_10}. The residuals are plotted as functions
of Galactic and ecliptic latitude, $b$ and $\beta$, in Figs.~\ref{gcr_b_grad}
and~\ref{gcr_beta_grad}. This model appears to do an excellent job at
removing the ISM emission variation 
all the way down to $b \sim 0\arcdeg$. The increase
in noise in the residual map occurs as a result of the introduction of the
noisy 140 $\micron$ data as part of the model. The noise partially obscures
some deficiencies in the residual map, such as some remaining emission from the
IPD which can be seen in a wide band along the ecliptic. This residual
emission arises primarily from the incomplete subtraction of IPD emission at 
100 $\micron$. 

Physically, the ISM2 model subtraction works well because it allows 
us to account for both the large-scale and local temperature variations of the 
interstellar dust. This is demonstrated by noting that the coefficients in
Equation~\ref{eq2color} define a line in the 100 -- 140 -- 240 $\micron$ 
color-color plot (Fig.~\ref{ccplot}). This line is a good
representation of the data, which are distributed along a nearly linear trend
because of temperature variations in the ISM. The bright source blanking
(Section 4.1) has removed the hottest sources (mostly H II regions) which
would extend the range of the data in the color-color plot, but with a
distinctly nonlinear trend. Only high signal-to-noise data 
($\nu I_{ZG_S} > 1200$ nW m$^{-2}$ sr$^{-1}$) were used 
in constructing Figure~\ref{ccplot}, therefore
small errors in the zero levels of each band will have little effect on the
colors. However, the gain uncertainties (\S6.1 and Table~\ref{tab_errors}) 
are large enough to cause systematic changes in the colors. Gain errors of 
$\sim7\%$ in each color would cause the
displacement between the observed colors and the
expected color trend for blackbodies with emissivities $\sim \nu^2$.

\subsection{Near-IR Intensities}

For the near-IR wavelengths, the subtraction of the starlight is not
sufficiently accurate to yield values of $R(\lambda)$ and $I_0(l,b,\lambda)$ 
from direct correlations of the near-IR data with $G_I(l,b,100$ $\micron)$.
The 1.25 -- 4.9 $\micron$ emission maps in Figure~\ref{gfs_map} show little
similarity to the 100 $\micron$ emission maps in Figure~\ref{zlsub_maps}.
Therefore, to determine the intensity of the ISM emission at these wavelengths
we make use of the color differences between the starlight and the ISM
emission, as revealed in the IPD-subtracted maps ($I_Z \equiv I_{obs} - Z$). 
Colors formed from the simple ratio of two near-IR intensities, e.g.
$I_Z(l,b,1.25$ $\micron) / I_Z(l,b,2.2$ $\micron)$, can reveal extinction
effects as well as intrinsic source variations. We remove the influence of
extinction by using the reddening-free parameter: 
\begin{equation}
Q(1.25,2.2,3.5) = [I_Z(1.25)/I_Z(2.2)]^{\zeta} / [I_Z(2.2)/I_Z(3.5)]
\end{equation}
where
\begin{equation}
\zeta = (A_{2.2}/A_{1.25} - A_{3.5}/A_{1.25}) / (1 - A_{2.2}/A_{1.25})
\end{equation}
and $A_{\lambda}/A_{1.25}$ is the ratio of extinction at wavelength $\lambda$ 
to that at 1.25
$\micron$. The ratio of absorption coefficients is calculated from the
extinction law of Rieke \& Lebofsky (1985), which has been shown to match the
inner Galaxy extinction seen by DIRBE (Arendt et al. 1994). 

The reddening-free parameter is constructed to be independent of the amount of
extinction; it only varies if the sources of emission vary. In fact,
$Q(1.25,2.2,3.5)$ is fairly insensitive to stellar spectral type, so the bulk
of the stellar emission of the Galaxy can be characterized by a single value,
$Q_0(1.25,2.2,3.5)$. Thus, variations in $Q(1.25,2.2,3.5)$ will be observed
where there is significant {\it emission} from hot interstellar dust. The map
of $Q(1.25,2.2,3.5)$ (in Figure~\ref{q123}) does show more structure of the
ISM at low latitudes than do the near-IR intensity maps of
Figure~\ref{gfs_map}, though residual IPD emission still obscures high
latitude structure. 

Based on the observation that the $I_Z(l,b,1.25$ $\micron) / I_Z(l,b,2.2$
$\micron)$ map shows little or no evidence of diffuse ISM features at $|b|
\gtrsim 10\arcdeg$, while the $I_Z(l,b,2.2$ $\micron) / I_Z(l,b,3.5$
$\micron)$ map shows such features clearly, we assume the ISM emission is only
significant at 3.5 $\micron$, i.e. 
\begin{equation}
I_Z(1.25) = G_S(1.25)
\end{equation}
\begin{equation}
I_Z(2.2) = G_S(2.2)
\end{equation}
and
\begin{equation}
I_Z(3.5) = G_S(3.5) + G_I(3.5).
\end{equation}
Then we can derive the ISM emission as 
\begin{equation}
G_I(3.5) = I_Z(3.5) \times [1 - Q_0(1.25,2.2,3.5)/Q(1.25,2.2,3.5)].
\end{equation}

Rather than imposing a value of $Q_0(1.25,2.2,3.5)$, we can again use a linear
least squares fit to derive $Q_0(1.25,2.2,3.5)$ and the near-IR cirrus color,
$R(3.5)$. We do this by considering 
\begin{eqnarray}
R(3.5)&\equiv&G_I(3.5) / G_I(100)\\
&=&I_Z(3.5) \times [1 - Q_0(1.25,2.2,3.5)/Q(1.25,2.2,3.5)] / G_I(100)\\
&=&\frac{Q(1.25,2.2,3.5) - Q_0(1.25,2.2,3.5)}{Q(1.25,2.2,3.5) \times G_I(100) / 
I_Z(3.5)}.
\end{eqnarray}

By rewriting this last equation as 
\begin{equation}
Q(1.25,2.2,3.5) = R(3.5) \times [Q(1.25,2.2,3.5) \times G_I(100) / I_Z(3.5)] + 
Q_0(1.25,2.2,3.5)
\end{equation}
we see that the color $R(3.5)$ can be found as the slope of a least-squares fit
of $Q(1.25,2.2,3.5)$ vs. $[Q(1.25,2.2,3.5) \times G_I(100) / I_Z(3.5)]$. The 
intercept of
the fit is $Q_0(1.25,2.2,3.5)$, the nominal stellar reddening-free parameter.
In a nearly identical manner we can also derive the $R(4.9)$ color of the ISM
emission. Finally, having determined these near-IR colors, we can use them 
to scale and subtract the 100 $\micron$ template of the ISM emission from the
3.5 and 4.9 $\micron$ data.

Because of the very low intensity of the ISM in the near-IR relative to the
residual errors of the IPD model, the correlations need to be done at low
Galactic latitudes ($|b|<30\arcdeg$). The correlations and corresponding
least-squares fits are shown in Figure~\ref{b3_b8}. The 3.5 and 4.9 $\micron$
colors derived by this method are listed in Table~\ref{tab_bx_b8}. 
The statistical uncertainties in the determination of the 3.5 and 4.9
$\micron$ colors are as good those for the far-IR wavelengths. The residual 
maps after subtraction of the derived near-IR emission are shown in 
Figure~\ref{gcr_maps}. (At 4.9 $\micron$, the residual
IPD emission is also reduced because the 100 $\micron$ ISM template contains
some residual IPD emission as well.)

In this derivation of the near-IR emission there are two important
assumptions. The first is that the near-IR emission of the ISM within the
high-latitude regions of interest is similar to that which is measured at low
latitudes by this procedure. The bright source blanking at low latitudes
removes most of the compact emission sources (mainly H II regions) in the 
area where the ISM colors are derived. The remaining emission may
still arise from ISM with different properties than those of the local
high-latitude ISM, but the residual maps (Figure~\ref{gcr_maps}) give no 
indication of differences between the unblanked low and high latitude ISM
emission. The second assumption is that any near-IR emission of the CIB can be 
neglected in this analysis. At the low galactic latitudes where the ISM colors
are derived, stellar emission is likely to be much stronger than the CIB 
emission. Additionally, the slopes of the correlations, which indicate the
ISM colors $R(3.5)$ and $R(4.9)$, should not be seriously affected by any
isotropic emission, such as that of the CIB.

\section{Uncertainties}

The value of the foreground subtracted maps (Figure~\ref{gcr_maps}) as
estimators of the CIB intensity depends upon of the uncertainties associated
with the data and foreground models. We distinguish two classes of
uncertainties. Random uncertainties, such as those arising from the detector
noise, are associated with errors exhibiting Gaussian or Poisson statistics.
In general, random uncertainties can be reduced by averaging large amounts of
data (i.e. over large regions of the sky, and over the entire mission) and do
not contribute significantly to the total uncertainty of the residual maps.
Systematic uncertainties, such as those associated with the detector offsets,
affect many or all data in a similar fashion and thus do not average out as
the size of the data set increases. The following sections describe our
estimates of the most significant systematic uncertainties associated with
each step of the foreground removal process. 

\subsection{Instrumental Uncertainties}

The calibration of the DIRBE data requires the measurement of instrument gain
and offset terms. Errors in the 
measurement of these quantities will affect all data in a similar fashion. The
evaluation of the gain and offset uncertainties is described in the {\it COBE}
DIRBE Explanatory Supplement (1997). The uncertainties in the detector gains
and offsets are listed in Table~\ref{tab_errors}. Gain errors will have little
impact on the significance of a detected background signal, because the IPD
and ISM models are scaled directly to the DIRBE data. The FSM is not scaled 
to the DIRBE data directly, but does use the DIRBE absolute calibration for
conversion from magnitudes to flux densities. A gain error will cause
us to overstate or understate the levels of the residual signal {\it and} its
total uncertainty by a multiplicative constant. The detector offset
uncertainties are much more important. They represent minimum uncertainties
that cannot be reduced. A true background signal must at least exceed the
offset uncertainties to be detected. The detector offset uncertainties are
worst for the 140 and 240 $\micron$ bands. 

After averaging over the entire mission, random instrumental noise is only
apparent in the 140 and 240 $\micron$ maps as a graininess visible at the
fainter high latitudes. However, averaging over regions larger than $\sim$ 30
pixels ($\approx 10^{-3}$ sr $\approx 3$ deg$^{2}$), reduces the uncertainty
from the instrumental noise to levels below that of the total systematic
uncertainties. At shorter wavelengths, the instrumental noise is dwarfed by
the systematic uncertainties even on the scale of a single pixel. 

\subsection{Uncertainties of the IPD Model}

The uncertainties and errors in the subtraction of the emission and scattered
light from the IPD are reported in Paper II. Of these uncertainties, the
one likely to be most significant is that arising from uncertainty in the 
geometry of the interplanetary dust cloud. The adopted geometric kernel does 
not uniquely or perfectly describe the data, and 
consequently we are required to associate
systematic errors to the residual intensities that are roughly proportional to
the mean intensity of the IPD at each wavelength. These uncertainties are
listed in Table~\ref{tab_errors}. 

\subsection{Uncertainties of the Faint Source Model}

The systematic uncertainty associated with the removal of Galactic stellar 
emission has two sources. First, there is uncertainty associated with
the bright source blanking. At some locations the blanking may 
remove confused sources which are individually below the bright source 
threshold, but are combined by the DIRBE beam into a single source above the 
threshold. The result of this effect is to reduce the mean level of the 
background when measured over large regions. We have used the Faint Source 
Model to estimate the frequency with which we expect to find pairs of sources
in the same DIRBE pixel, where both sources are within a factor of 2 of the
bright source threshold. The number of such double sources at latitudes
$|b|>30\arcdeg$ ranges from $\sim$ 140 at 1.25 $\micron$ to $\sim$ 6 at 4.9
$\micron$. The uncertainties caused by these sources are listed in
Table~\ref{tab_errors}.

A more important source of systematic uncertainties in the removal of stellar
foreground emission is the accuracy of the Faint Source Model. Systematic
errors in the Faint Source Model are not strictly isotropic, but occur on
large angular scales such that averages over large patches usually do not
reduce the errors. The clearest indication of systematic errors in the Faint
Source Model is the presence of gradients with respect to Galactic latitude in
the residual emission at 1.25 to 4.9 $\micron$. The correlations between these 
gradients and the FSM at $|b|\gtrsim30\arcdeg$ are used to derive the
uncertainties reported in Table~\ref{tab_errors}. The gradients indicate that
low latitude emission is oversubtracted relative to the high latitude
emission, but they cannot indicate at which latitude (if any) the subtraction
is correct. 

\subsection{Uncertainties of the ISM Model}

We have estimated two contributions to the uncertainty associated with the
subtraction of the ISM. The first is the uncertainty of the background level
that is removed from the 100 $\micron$ map to make the ISM template (see
Sections 5.2, 5.3). Any error in this value is propagated into the residual
intensities measured at other wavelengths. The uncertainty in the 100 $\micron$
background intensity is taken to be the difference in the intercepts
of the correlations between the H I and 100 $\micron$ emission at the NEP and
at the Lockman Hole (Section 5.2.1),
which were used to derive the 100 $\micron$ background intensity. 
This uncertainty would be larger if we had determined the background using 
a larger region of the sky, where there would be stronger and less well known 
contibutions from molecular and ionized components of the ISM. 
The 100 $\micron$ instrumental offset and
IPD uncertainties are added in quadrature to this ISM uncertainty before the
value is propagated to other wavelengths. The other ISM uncertainty is that
associated 
with the scaling factors, $R(\lambda)$, used to convert the 100 $\micron$ ISM
emission into ISM emission maps at other wavelengths. These uncertainties were
estimated by subdividing the region of sky used to determine $R(\lambda)$ into
$\sim$ 18 regions, and then calculating $R(\lambda)$ for each subregion. The
rms variation of $R(\lambda)$ for the set of subregions was used as the
uncertainty for the global $R(\lambda)$ applied in the ISM subtraction. This
uncertainty characterizes the variations in the ISM spectrum on angular scales
of $\sim30\arcdeg$. Both of the systematic uncertainties associated with the
ISM subtraction are listed in Table~\ref{tab_errors}. 

\subsection{Total Uncertainties of the Residual Emission}

After removal of the IPD and Galactic IR foregrounds, the total uncertainty
attributed to the residual background intensity at each wavelength is
calculated as the quadrature sum of the systematic uncertainties
($\sigma_{\rm total} = \sqrt{\Sigma\sigma_i^2}$) discussed above and listed in
Table~\ref{tab_errors}. 

\section{Interpretations}

\subsection{Modifications to the Faint Source Model}

In constructing the FSM, we attempted to reproduce the model described by 
Wainscoat et al. (1992) as closely as possible. This preserves the model's
basis in source counts and independence from the actual intensity of the CIB.
The only modification made to 
the model in order to improve its fit to the DIRBE data was the addition of an
18 pc offset of the Sun above the Galactic midplane. This value is supported by
the independent results derived at near-IR wavelengths by Cohen (1995), and
other analyses of the DIRBE data (Weiland et al. 1994; Freudenreich 1996)
bracket the value used here. However, the final residual maps in the near-IR
bands (Fig.~\ref{gcr_maps}), and the gradients with respect to Galactic
latitude (Table~\ref{tab_gcr_grad}) clearly indicate that the FSM has
overaccounted for the stellar emission. 

This suggests the possibility of adjusting the FSM to minimize these 
gradients. The simplest adjustment would be to apply a global 
scaling factor to the model intensities. Such a factor could arise if there 
were an error in the flux attributed to 0--magnitude stars in the FSM. The
scaling factors required to minimize the Galactic latitude gradients of the 
residual maps are 0.859, 0.792, and 0.674 at 1.25, 2.2, and 3.5
$\micron$, respectively. However, these adjustments are significantly 
larger than the $\sim 3\%$ gain uncertainties expected at these wavelengths
({\it COBE} DIRBE Explanatory Supplement, 1997).
These scaling factors could also be attributed to
errors in the absolute magnitudes of 0.165, 0.253, and 0.428 magnitudes for
all source types at 1.25, 2.2, and 3.5 $\micron$, respectively. If these
scaling factors are applied to the FSM, then the residual intensities at high
Galactic latitude increase by factors of $\sim$2 in all three bands. 

Alternatively, we can use the star count data from the seven 2MASS fields to 
optimize the fit of the FSM to this independent data set. To minimize the 
$\chi^2$ statistic for the comparison of the $J$ and $K$ band data 
simultaneously, we find that the FSM star counts (number of stars/mag/deg$^2$) 
need to be reduced by a factor of 0.952 to match the 2MASS data. Such a change 
in the number density of stars in the FSM would eliminate about 1/2 of the 
gradient that is observed in the residual maps. Changes that optimize the fit 
to the 2MASS star counts do not simultaneously minimize the residual gradients 
in the DIRBE data.

None of these scaling changes were applied to either the intensities or the
number densities of the stars in the FSM in the present analysis because we
sought to keep the FSM tied closely to the star counts, and because we are
unable to determine the physical cause of the apparent discrepancy between 
the FSM and the DIRBE data. 

\subsection{Implications for the ISM}

The IR emissivity and colors of the ISM derived in this paper differ somewhat
from those found in other analyses. Using the DIRBE data, but a different
method for removing the IPD emission, Boulanger et al. (1996) find a 100
$\micron$ emissivity of 15.9 nW m$^{-2}$ sr$^{-1}$ / 10$^{20}$ cm$^{-2}$, 
which is about 15\% lower than the average for the NEP and Lockman Hole
regions (Table~\ref{tab_b8_hi}). Dwek et al. (1997) presented an analysis of
an ISM spectrum that was derived in a similar manner to that presented in
Table~\ref{tab_bx_b8}, but using an earlier version of the IPD emission
removal than that used in this work (Paper II). The high latitude ISM spectrum
presented here is considered an improvement on that presented by Dwek et al.
(1997), though differences are generally $\lesssim$ 10\% and within the
adopted uncertainties. 

Dwek et al. (1997) find that the ISM spectrum is well fit by emission from
silicate grains, graphite grains, and PAHs, all heated by the mean
interstellar radiation field. The near-IR emission ($\lambda \leq 12~\micron$)
provides strong evidence that the PAH abundance is higher than expected from
most previous studies. The amount of carbon in graphite grains, PAHs, and as
C$^+$ ions is found to be consistent with cosmic abundances. 

At 240 $\micron$ the results of the two-component ISM model (ISM2) indicate 
that it is important for a model of Galactic IR emission to account for 
spatial variation in the dust temperature. The contrast between 
Figure~\ref{gcr2_10} and Figure~\ref{gcr_maps} shows this 
is especially relevant at low latitude ($|b| \lesssim 20\arcdeg$). However,
because of the increased noise in the ISM2 model, it remains unclear how
important temperature variation is at high latitudes. A two-component model,
similar in form to the ISM2 model used here, is also required to model the
Galactic emission observed by the {\it COBE}/FIRAS experiment (Fixsen et al.
1997). Another recent study using both DIRBE and FIRAS data (Lagache et al. 
1998) decomposes the ISM emission into warm ($\sim$17.5 K) and colder 
($\sim$15 K) components. The distribution of this cold component (Figure 5
of Lagache et al. 1998) is very similar to that of the 140 and 240 $\micron$ 
residual emission after removal of the one-component ISM model (Figure 12).
In this two temperature model, the apparent temperature variations revealed 
in the long wavelength colors are (at least in part) a result of varying 
proportions of the warm and colder ISM components on different lines of sight.

\subsection{Inadvisable Estimates of the CIB}

There are several numbers presented in this paper that should not be used as 
estimates of the CIB, despite their appearances. These include extrapolation 
of the $\csc(|b|)$ or $\csc(|\beta|)$ gradients discussed in Sections 4.3 and 
5.3, intercepts of the correlations between ${\nu}I_{ZG_S}$(100 $\micron$) and 
$N$(H I) (and other ISM tracers) discussed in Section 5.2, and the intercepts
of the correlations of other DIRBE intensity maps with the 100 $\micron$ ISM
template discussed in Section 5.3. The main reason these numbers are not
reliable estimates of the CIB is that they are derived from regions chosen for
characterization of the Galactic foreground, not for accuracy or completeness
of removal of the Galactic and other foregrounds. Frequently these regions
include entirely or in part areas where the Galactic emission is relatively
strong, and small fractional errors in the foreground removal could still have
large influences on the derived intensity of the CIB. A more useful 
investigation of the CIB should focus on regions of the sky where errors in
the foreground removal have minimal impact on the derived CIB. A proper
analysis of the CIB also needs a demonstration that the residual emission is
isotropic, rather than just providing a single intensity estimate that may
average over anisotropic defects in the foreground removal. Such a study of the
CIB is presented in Paper I. Any conclusions concerning the CIB
should be drawn from that paper. 

\section{Conclusions}

We have modeled and removed the Galactic IR emission in the DIRBE data in 
preparation for analysis of the CIB. The procedures used were designed to 
preserve the emission of the CIB in the residual maps, and not remove it 
inadvertently with the Galactic emission. The procedures concentrated on 
producing accurate results at high latitudes where Galactic emission is
weakest. At low latitudes, deficiencies in the models are clearly visible in
the residual maps. 

We find that the stellar emission of the Galaxy is reasonably reproduced by 
our Faint Source Model, which is based on the SKY model (Wainscoat et al. 1992; 
Cohen 1993a, 1994a, 1995). An offset of the Sun by $\sim$ 18 pc
from the Galactic plane is required to produce equal residual near-IR
intensities at north and south Galactic latitudes. However, there is
clearly room for improvement in the geometry or calibration of the FSM. 

We find that the ISM can be fairly well modeled by a single spatial and
spectral component if we constrain our study to high Galactic latitudes. At
240 $\micron$, a model of the ISM with two spatial components can produce a
much more complete subtraction of the ISM, extending to low Galactic
latitudes. The two spatial components can combine to produce a range of color
temperatures across the sky. This shows that a complete model of the ISM needs
to be able to account for a continuous range of dust temperatures. We are 
unable to detect any IR emission associated with low density ionized gas at
high Galactic latitudes. 

\acknowledgments

The authors gratefully acknowledge the contributions over many years of the 
many engineers, managers, scientists, analysts, and programmers engaged in the 
DIRBE investigation. The National Aeronautics and Space Administration/Goddard 
Space Flight Center (NASA/GSFC) is responsible for the design, development, 
and operation of {\it COBE}. Scientific guidance is provided by the {\it COBE} 
Science Working Group. GSFC is also responsible for the development of the 
analysis software and for the production of the mission data sets.

H. T. Freudenreich is thanked for his work on developing and implementing the 
procedures used for blanking the bright sources (Section 4.1). We thank M.
Cohen, who provided results from the SKY model and technical assistance, which
were very useful for the development of our Faint Source Model. We also thank
M. Skrutskie, S. Price, R. Cutri, M. Egan, and others in the 2MASS project for
making some of their prototype camera data available for this research. 

The data described in this paper are available to the public through the NSSDC 
{\it COBE} homepage website at 
${\tt http://www.gsfc.nasa.gov/astro/cobe/cobe\_home.html}$. 
The mission-averaged zodi-subtracted residual skymaps are contained in the 
`Zodi-Subtracted Mission Average (ZSMA)' maps; the results generated by the 
Faint Source Model are contained in the `Faint Source Model (FSM)' maps.

\appendix

\section*{Appendix: Far-Infrared Emissivity of the Diffuse Ionized Medium}

In Section 5.2.2, tracers of ionized gas toward the Lockman hole region 
were used to estimate the possible error in the zero level of the 
100 $\mu$m ISM template, as determined from intercepts of
$\nu I_{ZG_S}$(100 $\mu$m) -- $N$(H I) correlations, 
that result from our neglect of emission 
from H II associated dust. The 100 $\mu$m emissivity per H nucleus was assumed
to be the same in the ionized gas and the neutral atomic gas. Here, an attempt 
is made to determine the mean emissivity per H nucleus for the ionized gas at 
high galactic latitudes. 

Previous efforts to measure far-infrared emission 
from the diffuse ionized medium have been inconclusive. Boulanger
et al. (1995) found that the extended, low density H II region around the high
latitude early B star $\alpha$ Vir has associated far-IR emission 
at a brightness that is consistent with a normal dust abundance, but this 
region may not be representative of the high latitude H II in general. 
Boulanger et al. (1996) searched for far-IR emission from the ionized medium by
looking for a $\csc(\vert b \vert)$ dependence in residuals of their high 
latitude far-IR--$N$(H I) fits, which used DIRBE and 
FIRAS data from 100 to 1100 
$\mu$m. From the slopes of their $\csc(\vert b \vert)$ fits, they placed 
an upper limit on the far-IR brightness of a component of the H II gas that is 
not spatially correlated with $N$(H I). They found this limit to be consistent 
with a normal dust abundance in the ionized medium if about half of the far-IR 
emission from the ionized medium is uncorrelated with $N$(H I). Determination 
of the emissivity per H nucleus in the H II gas by 
this approach would require knowledge of the latitude dependence of H$_2$ 
associated emission, the latitude dependence of emissivity per H nucleus in 
the H I gas, and the degree of correlation of $N$(H II) with $N$(H I), which
may also vary with latitude. 
Also using the FIRAS data, Fixsen et al. (1998) modeled the high latitude ISM 
emission with a combination of 158 $\micron$ [C II] line emission, 
a tracer of ionized gas, and linear and 
quadratic terms in $N$(H I), a tracer of the neutral gas. Their analysis found
that most of the high latitude far-IR emission correlated with the quadratic 
term of the H I column density, and essentially no continuum emission 
correlated with the [C II] line emission. 
There have been a number of studies of
correlations between far-IR and H$\alpha$ emission or 
between far-IR and microwave
(largely free-free) emission at high latitudes (e.g., Kogut et al. 1996a,
1996b; De Oliveira-Costa et al. 1997; McCullough 1997; Kogut 1997).
Interpretation of these correlations is not straightforward because they do
not separately account for the far-IR emission from the 
dominant H I phase of the
interstellar gas. Sodroski et al. (1997) derived mean 
far-IR emissivities for the
extended low density ionized gas at low galactic latitudes by decomposing
DIRBE maps into components that correlate with H I, CO, and radio continuum
emission. This gas is probably associated with star forming regions, and is
probably much denser and subject to a stronger radiation field than the
diffuse warm ionized medium observed at high latitudes (e.g., Heiles, Reach,
and Koo 1996; Lockman, Pisano, and Howard 1996). 

Here, we use observations of H$\alpha$ intensity at high latitudes and 
observations of dispersion measure of pulsars at high $\vert z \vert$ as 
tracers of
the ionized gas. We compare results of regression fits of the forms
$\nu I_{ZG_S}(100$ $\micron)$ = $A~N$(H I) + $B$ and $\nu I_{ZG_S}(100$ 
$\micron)$ = $C~N$(H I) + $D~f$(H II) + $E$
for data along high latitude lines of sight, where $I_{ZG_S}(100$ $\micron)$ is 
DIRBE intensity after subtraction of IPD emission and $f$(H II) is either
extinction-corrected H$\alpha$ intensity or pulsar dispersion measure. The 
derived value of the parameter $D$ can provide an estimate of the 100 $\mu$m 
emissivity 
per H nucleus of the ionized medium, and the difference between parameters 
$B$ and $E$ provides an estimate of error in the zero level of 100 $\mu$m ISM 
emission that is inferred if the ionized component is not accounted for.

Each tracer of ionized gas has advantages and disadvantages.
Pulsar dispersion measure is a measure of ionized gas column density,
but it pertains to a single line of sight rather than the solid angle
sampled by the DIRBE beam, and it can significantly underestimate the total 
ionized
column density along the line of sight if the pulsar is not above most of the 
ionized gas layer. The H$\alpha$ data used here were measured with a beam 
comparable to the DIRBE beam, and sample the entire path through the galaxy. 
However, H$\alpha$ emissivity is proportional to the square of the ionized 
gas density while 100 $\mu$m emissivity is proportional to dust density.
Another disadvantage of using H$\alpha$ is that it can be affected by
extinction. An extinction correction has been applied to the H$\alpha$
intensities used here assuming the ratio of extinction coefficient to 
H$\alpha$ emissivity is constant along each line of sight (uniformly mixed
extinction and emission), and the optical depth at H$\alpha$
is 0.04 $N$(H I) where $N$(H I) is in $10^{20}$ cm$^{-2}$. With these
assumptions, the extinction correction factors range from 1.02 to 1.24 
for the lines of sight used here.

The regression fitting has been done independently for five different regions
or samples of positions. Sample 1 consists of 95 positions in a 
$10\arcdeg\times12\arcdeg$ 
region centered at $l=144\arcdeg$, $b=-21\arcdeg$, which was 
mapped in H$\alpha$ 
with a $0\fdg8$ beam by Reynolds
(1980). The region contains a number of elongated H$\alpha$ enhancements above 
a smooth background, and positions toward the brightest feature were excluded 
from our analysis because its estimated electron density is about 5 times 
higher than that of the other features (Reynolds et al. 1995). Sample 2 is a 
sample of 27 positions within 20 degrees of $\alpha$ Vir (at $l=317\arcdeg$, 
$b=50\arcdeg$) that were observed in H$\alpha$ by Reynolds (1985).
Two positions within 1 degree of $\alpha$ Vir were not included in this sample
because an
enhanced 60 $\mu$m to 100 $\mu$m intensity ratio is observed at these positions.
Sample 3 is a sample of 22 pulsar lines of sight at $\vert b \vert > 30\arcdeg$
that were observed in H$\alpha$ by Reynolds (1984, 1991b). 

Samples 4 and 5 are different samples of pulsar lines of sight at 
$\vert b \vert > 20\arcdeg$, for which pulsar dispersion measure
is used instead of H$\alpha$ as the tracer of ionized gas. Two different 
criteria were used to select pulsars likely to be at high $\vert z \vert$ 
distances, so the dispersion measure traces most of the ionized gas layer. 
The first criterion is based on estimates of the $z$ component of the 
characteristic pathlength $L_c$ occupied by 
ionized gas between the Sun and the pulsar, $L_c \equiv 
(\int^d_0 n_e ds)^2/\int^d_0 n_e^2 ds$, where $d$ is the distance of the 
pulsar. Following an approach similar to that of Reynolds (1977, 1991b), 
lower limits to $L_c$ were calculated for lines of sight to pulsars observed
in H$\alpha$ by Reynolds (1984, 1991b), using $L_c\, > \, DM^2/EM$. 
Here $DM$ is the pulsar dispersion measure and $EM$ is the emission measure
along the entire line of sight through the galactic disk, determined from
the extinction-corrected H$\alpha$ intensity assuming an electron temperature
of 8000 K. 
The calculated lower limit on $L_c \vert \sin\, b \vert$ ranges from 200 to 390 
pc for lines of sight toward pulsars in four globular clusters at $\vert z 
\vert > $ 4 kpc, and it is assumed that other pulsars with $L_c\, 
\vert \sin\, b \vert\, \ge$ 200 pc are also above most of the ionized gas. 
Sample 4 consists of the 9 pulsar lines of sight selected by this criterion.
Data for this sample are listed in Table~\ref{tab_pulsars}. 
Sample 5 consists of 22 pulsar 
lines of sight at $\vert b \vert > 20\arcdeg$ for which $DM$ $\vert\sin b\vert$ 
is greater than $0.6 \times 10^{20}$ cm$^{-2}$, the minimum value of 
$DM$ $\vert\sin b\vert$ for the globular cluster pulsars in 
Table~\ref{tab_pulsars}. 
Some of the pulsars selected by this criterion may not be above most of the 
ionized gas layer if density enhancements in the ionized medium are common 
at high latitudes. The enhancements observed by Reynolds et al. (1995) 
in the $10\arcdeg \times 12\arcdeg$ region at $l=144\arcdeg$, 
$b=-21\arcdeg$ have estimated 
$N$(H II) values ranging from 0.2 to 0.6 $\times 10^{20}$ cm$^{-2}$.

Results of the regression fits for Samples 1--3 are shown in 
Table~\ref{tab_hi_ha} and
results for Samples 4 and 5 are shown in Table~\ref{tab_hi_dm}. 
The errors listed were
determined from the size of the 68\% joint confidence region in parameter
space. The fits for Samples 1 and 2 used H I column densities from the 
Leiden/Dwingeloo survey (Hartmann and Burton 1997), smoothed
to the resolution of the DIRBE data, and the fits for the other samples used
H~I column densities from the Bell Laboratories survey (Stark et al. 1992).
Except for the 
region around $\alpha$ Vir, the derived coefficient of the ionized gas tracer 
is consistent with zero for each sample.
(For comparison, the mean 100 $\mu$m emissivities derived by Sodroski et al. 
(1997) for the extended low density ionized gas at low galactic latitudes 
correspond to $D_1$ = $2.6 \pm 0.1$ nW m$^{-2}$ sr$^{-1}$/Rayleigh inside the 
solar circle and $1.8 \pm 0.2$ nW m$^{-2}$ sr$^{-1}$/Rayleigh outside the 
solar circle, assuming an ionized gas temperature of 8000 K.)
The derived values of the 
isotropic terms $B$ and $E$ agree with each other within the uncertainties, and
they are also consistent with the value of 19.8 nW m$^{-2}$ sr$^{-1}$ used for
making the 100 $\mu$m ISM template (Section 5.3). 

Figure~\ref{sample_corrs} illustrates correlations among the variables used
for the regression fit for Samples 1, 2, and 4. Figure~\ref{sample_corrs}a
shows linear fits to $\nu I_{ZG_S}$(100 $\mu$m)--$N$(H I) correlations,
Figure~\ref{sample_corrs}b shows linear fits to H$\alpha$--$N$(H I) or $N$(H
II)--$N$(H I) correlations, and Figure~\ref{sample_corrs}c shows the
correlation between residuals of the fit in Figure~\ref{sample_corrs}a and
residuals of the fit in Figure~\ref{sample_corrs}b. Significant trends are
not seen in Figure~\ref{sample_corrs}c, except for the region around $\alpha$
Vir. 

The results of the regression fits and of Figure~\ref{sample_corrs} show that
100 $\mu$m emission has been detected from the ionized region around $\alpha$
Vir, but has not been detected from the general ionized medium at high
latitudes. 
For Sample 1, the three sigma
upper limit of $D_1 \, < \, 4.2$ nW m$^{-2}$ sr$^{-1}$/Rayleigh corresponds to
an upper limit on the 100 $\mu$m emissivity per H nucleus in the ionized
medium of $D_2 < 12$ nW m$^{-2}$ sr$^{-1}$/10$^{20}$ cm$^{-2}$ , assuming
electron density $n_e \le 0.2$ cm$^{-3}$ and electron temperature of 8000 K
(Reynolds et al. 1995). The results for Sample 3 give $D_2 \, < \, 4$ nW
m$^{-2}$ sr$^{-1}$/10$^{20}$ cm$^{-2}$ under the same assumptions. Comparison
with the values of C found for these samples suggests that the 100 $\mu$m
emissivity per H nucleus is smaller in the ionized medium than in the neutral
atomic medium. The results of the fits for Samples 4 and 5 are consistent
with this conclusion, but the errors in the derived values of $D_2$ are large.

Differences in the electron density distribution along different lines of 
sight may cause errors in the results of the regression fits that use H$\alpha$ 
intensity as the ionized gas tracer. For the region of Sample 1, a crude 
model of the electron density distribution has been constructed.
For each of the elongated features of enhanced H$\alpha$ emission, we use
values derived by Reynolds et al. (1995) for the mean electron density 
and the mean linear size along the line of sight. 
The densities range from 0.08 to 0.23 cm$^{-3}$. 
For the gas emitting the 
smooth H$\alpha$ background, we assume an electron density of 
0.08 cm$^{-3}$ (Reynolds 1991b) 
and determine its extent along each line of sight
from the observed H$\alpha$ intensity, allowing for the 
contribution of each enhancement and assuming a gas 
temperature of 8000 K.
The regression fit for Sample 1 was then repeated using ionized column density
from the model instead of H$\alpha$ intensity. Results are $C=16.0 \pm 1.1$
nW m$^{-2}$ sr$^{-1}$/10$^{20}$ cm$^{-2}$, $D_2= 0.6 \pm 1.9$ nW m$^{-2}$ 
sr$^{-1}$/10$^{20}$ cm$^{-2}$, and $E=26.1 \pm 8.3$ nW m$^{-2}$ sr$^{-1}$. 
The electron density model
is not unique, but the results are consistent with the conclusions that
the 100 $\mu$m emissivity per H nucleus is smaller in the ionized medium
than in the neutral atomic medium, and that 100 $\mu$m emission from the ionized
medium does not cause significant error in the zero level of 100 $\mu$m ISM
emission inferred from the $\nu I_{ZG_S}$(100 $\mu$m) -- $N$(H I) correlation. 

Errors in the fit parameters can also occur if there is significant 100 $\mu$m
emission from H$_2$-associated dust along one or more lines of sight in a
sample, but for the samples used here it appears that such errors are small.
The regression fits were repeated excluding positions in each sample where 
the 100 $\mu$m emission exceeds the H I correlated component of 100 $\mu$m
emission by more than about 20 nW m$^{-2}$ sr$^{-1}$. Reach et al. (1994, 
1997) have presented evidence that such excess far-IR emission generally 
traces molecular gas.
A small fraction of the positions was excluded for each sample, and the
results were not significantly different from those of 
Table~\ref{tab_hi_ha} and Table~\ref{tab_hi_dm}.

Other possible sources of error include errors in subtraction of IPD emission,
and possible differences in the mean 100 $\mu$m emissivity per H nucleus, 
within either the H I or H II gas, for different lines of sight. Such 
differences could be caused by differences in mean dust temperature or mean 
dust-to-gas mass ratio. Use of a small patch of the sky with a large number 
of lines of sight may reduce these sources of error, so the results for
Sample 1 are considered to be the most reliable.
 
To summarize, a multiple regression analysis has been used to estimate an 
upper limit on the 100 $\mu$m emissivity per H nucleus in the diffuse ionized 
gas at high latitudes. An upper limit of 12 nW m$^{-2}$ 
sr$^{-1}$/10$^{20}$ cm$^{-2}$ is obtained for a $10\arcdeg \times 12\arcdeg$ 
region at $l=144\arcdeg$, $b=-21\arcdeg$. This value is 3/4 of the emissivity
per H nucleus in the neutral atomic gas in the same region. Results for other
samples of high latitude positions are consistent with this result, except a
greater 100 $\micron$ emissivity is found for ionized gas in the vicinity of
the early B star $\alpha$ Vir. 
If our derived upper limit on the H II/H I emissivity ratio is valid for the 
general high latitude sky, the mean column density ratio $N$(H II)/$N$(H I) of
about 1/3 (Reynolds 1991a) implies that on average less than $\sim$20\% of 
the total 100 $\micron$ emission observed at high latitudes comes from the 
ionized gas phase.
A low far-IR emissivity per H nucleus might be
expected for the diffuse ionized medium since dust destruction by shocks is
thought to be more efficient in the warm diffuse interstellar gas than in
dense clouds (e.g., Seab 1987; McKee 1989; Jones et al. 1994, 1996). 
Future results from the Wisconsin H-Alpha Mapper survey (Tufte et al. 1996)
will be useful for extending the type of analysis presented here to large 
areas of the high latitude sky.

\newpage

\figcaption[]{Full sky DIRBE intensity maps in Galactic Mollweide projection. 
Only IPD scattered light and emission have been removed. Intensities are 
scaled linearly in the ranges ($-0.05$,0.3), ($-0.05$,0.3), ($-0.01$,0.2), 
(0,0.2), (0,2), (0.5,3), (0,3), (0,15), (0,20), (0,20) MJy/sr at wavelengths 
from 1.25 to 240 $\micron$. Any intensities above or below these ranges 
appear white or black, respectively.
\label{zlsub_maps}}

\figcaption[]{The 2.2 $\micron$ intensity map after application of bright 
source blanking. Intensity range is the same as in Figure 1.
\label{bs_map}}

\figcaption[]{The 2.2 $\micron$ intensity map of the Faint Source Model. 
Intensity range is the same as in Figure 1.
\label{fsm_map}}

\figcaption[]{1.25 -- 4.9 $\micron$ intensity maps after removal of emission 
from the IPD, and bright and faint Galactic stellar sources. Intensity ranges 
are the same as in Figure 1.
\label{gfs_map}}

\figcaption[]{Galactic intensity gradients. The left column shows the 
intensity after the IPD emission and bright sources have been removed as a
function of $\csc(|b|)$. The right column shows the same data after the
Faint Source Model has also been subtracted. [$\csc(15\arcdeg)
\approx 4, \csc(30\arcdeg) = 2.0,$ and $\csc(60\arcdeg) = 1.15$].
\label{gfs_b_grad}}

\figcaption[]{Ecliptic intensity gradients. Same as Fig. 5, but plotted with
respect to ecliptic rather than Galactic latitude. 
\label{gfs_beta_grad}}

\figcaption[]{Correlation of DIRBE 100 $\micron$ intensities after subtraction
of IPD emission with H I column density for (a) the North
Ecliptic Pole region, (b) a region of low H I column density in Ursa Major,
and (c) the region of high-latitude sky ($\vert{\it b}\vert$ $>$ 25\arcdeg,
$\vert{\it \beta}\vert$ $>$ 25\arcdeg) covered by the Bell Laboratories H I
survey. The solid lines show fits to the data as described in \S 5.2.1. Data
above the short diagonal lines were excluded from the fitting. 
\label{b10_hi}}

\figcaption[]{Correlations of Figure 7 averaged within bins along the fit
lines. The bin boundaries used are perpendicular to the fit lines on plots
where each variable is divided by its mean measurement error. The error bars
are parallel to the bin boundaries and show the standard error of the mean for
the data within each bin.
\label{b10_hi_binned}}

\figcaption[]{(a) Ecliptic latitude dependence of DIRBE 100 $\micron$ data
after subtraction of IPD emission. Data at $\delta > -40\arcdeg$ are
averaged within 10 degree bins in ecliptic latitude and the following bins in
H I column density: $N$(H I)/10$^{20}$ cm$^{-2}$ from 1.3 to 1.5 (diamonds), 1.5
to 1.7 (crosses), 1.7 to 1.9 (triangles), and 1.9 to 2.1 (squares). (b) The
data averaged within 10 degree bins in Galactic latitude and the same bins in
H I column density. (c) The data averaged within 20 degree bins 
in Galactic longitude and the same bins in H I column density.
\label{lat_trends}}

\figcaption[]{The ratio of H$\alpha$ intensity to dispersion
measure as a function of H I column density, for lines of sight toward pulsars
at $\vert b\vert > 20 \arcdeg$ for which the $z$ component of the characteristic
pathlength occupied by ionized gas, $L_c \vert \sin b\vert$, is 200 pc or
greater (see Appendix). These pulsars are probably above most of the
ionized gas layer, and each dispersion measure is probably close to the total
$N$(H II) along the line of sight. Squares show data toward pulsars that are in
globular clusters at $\vert z\vert > $4 kpc.
\label{ha_dm}} 

\figcaption[]{ISM correlation plots. These plots show the correlations that 
were used to derive the $R(\lambda)$ ISM colors at 12, 25, 60, 140, and 240 
$\micron$. The lines indicate the weighted least-squares fit to the data.
\label{bx_b8}}

\figcaption[]{3.5 -- 240 $\micron$ intensity maps after removal of all IPD and
Galactic foreground emission. Intensity ranges are the same as in Figure 1.
\label{gcr_maps}}

\figcaption[]{Galactic intensity gradients. The left column shows the intensity
after the IPD model, bright sources, and the Faint Source Model have been
removed as a function of $\csc(|b|)$. The right column shows the same
data after the ISM emission has also been subtracted.
\label{gcr_b_grad}}

\figcaption[]{Ecliptic intensity gradients. Same as Fig. 13, but plotted 
with respect to ecliptic rather than Galactic latitude.
\label{gcr_beta_grad}}

\figcaption[]{The 240 $\micron$ intensity map after removal of all foregrounds 
using the two-component ISM model (ISM2). Intensity range is the same as in 
Figure 1.
\label{gcr2_10}}

\figcaption[]{Far-IR color-color plot. Each point represents the colors of a 
pixel where the 100 $\micron$ brightness is $>$ 40 MJy/sr. The straight line is 
the trend implied by the coefficients of the two-component ISM model (Equation 
8). The crosses connected by lines indicate the expected
colors for sources with emissivities $\sim\nu^2$ and temperatures from 16 to
22 K.
\label{ccplot}}

\figcaption[]{The map of the reddening-free parameter, $Q(1.25,2.2,3.5)$.
Mid-latitude structures near $(l,b) = (90\arcdeg,+20\arcdeg)$ and
$(270\arcdeg,-20\arcdeg)$ can be seen to correlate with ISM emission (e.g. the
240 $\micron$ emission in Figure 1). Residual IPD errors cause artifacts
(extended blue regions) at low ecliptic latitudes. 
\label{q123}}

\figcaption[]{Near-IR ISM correlation plots. These plots show the correlations 
that were used to derive the $R(\lambda)$ ISM colors at 3.5 and 4.9 $\micron$. 
The lines indicate least-squares fits to the data.
\label{b3_b8}}

\figcaption[]{Correlation plots of data for the lines of sight in Samples
1, 2, and 4. The error bars show 1$\sigma$ measurement errors. The line
shown in each panel is an unweighted least squares fit that minimizes the
residuals in the ordinate.
(a) Correlation of 100 $\micron$ data with H I column density. (b) Correlation
of extinction-corrected H$\alpha$ intensity (for Samples 1 and 2) or H II
column density (for Sample 4) with H I column density. (c) Correlation of
residuals from the fit of (a) with residuals from the fit of (b).
\label{sample_corrs}} 

\begin{deluxetable}{cccc}
\tablecaption{Effects of Faint Source Subtraction on Galactic Intensity 
Gradients\label{tab_fsm_grad}}
\tablehead{
\colhead{Wavelength}&
\colhead{Before/After Faint}&
\colhead{Gradient\tablenotemark{a}}&
\colhead{Correlation} \\
\colhead{($\micron$)}&
\colhead{Source Subtraction}&
\colhead{(nW m$^{-2}$ sr$^{-1}$ / $\csc(|b|)$)}&
\colhead{Coefficient\tablenotemark{a}}
}
\startdata
1.25 & before & $192.7 \pm 0.7$ &  0.61\nl
1.25 & after  & $-21.6 \pm 0.7$ & -0.09\nl
2.2  & before & $ 74.3 \pm 0.3$ &  0.56\nl
2.2  & after  & $-16.8 \pm 0.3$ & -0.15\nl
3.5  & before & $ 29.4 \pm 0.1$ &  0.45\nl
3.5  & after  & $ -4.6 \pm 0.1$ & -0.07\nl
4.9  & before & $ 11.4 \pm 0.1$ &  0.24\nl
4.9  & after  & $ -0.4 \pm 0.1$ & -0.005\nl
\enddata
\tablenotetext{a}{$|b|>30\arcdeg$}
\end{deluxetable}

\begin{deluxetable}{cccccc}
\scriptsize
\tablecaption{Parameters of 100 $\micron$ -- H I Correlations\label{tab_b8_hi}}
\tablehead{
\colhead{}&
\colhead{Center of}&
\colhead{$\nu I_0(100$ $\micron)$}&
\colhead{$A$}&
\colhead{}&
\colhead{Correlation}\\
\colhead{Region}&
\colhead{Region}&
\colhead{(nW m$^{-2}$ sr$^{-1}$)}&
\colhead{(nW m$^{-2}$ sr$^{-1}$ / $10^{20}$ cm$^{-2}$)}&
\colhead{$\chi^2_\nu$}&
\colhead{Coefficient}
}
\startdata
North         & $(l,b) = (96,30)$          & $17.3 \pm 1.7$ & $19.1 \pm 0.5$ &  3.3 & 0.95\nl
Ecliptic Pole & $(\lambda,\beta) = (0,90)$ &  &  &  \nl
\nl
Lockman  & $(l,b) = (150,53)$           & $22.3 \pm 0.5$ & $17.6 \pm 0.5$ &  4.4 & 0.75\nl
Hole     & $(\lambda,\beta) = (136,45)$ &  &  &  \nl
\nl
$|b|> 25,$ &  & $21.2 \pm 0.6$ & $18.6 \pm 0.3$ & 12.9 & 0.81\nl
$|\beta|> 25, \delta > -40$  &  &  &  &  \nl
\enddata
\end{deluxetable}

\begin{table}
\dummytable\label{tab_hi_h2}
\end{table}

\begin{deluxetable}{ccccc}
\footnotesize
\tablecaption{Parameters of IR Correlations\label{tab_bx_b8}}
\tablehead{
\colhead{Wavelength}&
\colhead{$R(\lambda)\tablenotemark{a}$}&
\colhead{Dispersion}&
\colhead{Correlation}&
\colhead{Region}\\
\colhead{($\micron$)}&
\colhead{}&
\colhead{(nW m$^{-2}$ sr$^{-1}$)}&
\colhead{Coefficient}&
\colhead{}
}
\startdata
  1.25 & \nodata& \nodata& \nodata& \nodata\nl
  2.2  & \nodata& \nodata& \nodata& \nodata\nl
  3.5  & $0.00183 \pm 0.00001$ & 17.6 & 0.79 & \tablenotemark{b}\nl
  4.9  & $0.00291 \pm 0.00003$ & 22.0 & 0.66 & \tablenotemark{b}\nl
 12    & $0.0462  \pm 0.0001$  & 15.2 & 0.94 & $b > 10\arcdeg, \beta > 70\arcdeg$\nl
 25    & $0.0480  \pm 0.0002$  &  8.5 & 0.93 & $b > 10\arcdeg, \beta > 70\arcdeg$\nl
 60    & $0.171   \pm 0.0003$  &  4.0 & 0.91 & $|b|> 30\arcdeg, |\beta|> 40\arcdeg$\nl
100    & $1.00$ & \nodata& \nodata& \nodata\nl
140    & $1.696   \pm 0.008$   & 57.2 & 0.52 & $|b|> 45\arcdeg$\nl
240    & $1.297   \pm 0.005$   & 19.4 & 0.62 & $|b|> 45\arcdeg$\nl
\enddata
\tablenotetext{a}{$= I_{\nu}(\lambda) / I_{\nu}(100$ $\micron)$}
\tablenotetext{b}{For 3.5 and 4.9 $\micron$ the regions used were 
$|b|<30\arcdeg, |\beta|>40\arcdeg$, and $54\arcdeg<l<138\arcdeg$ and 
$234\arcdeg<l<318\arcdeg$.} 
\end{deluxetable}

\begin{deluxetable}{cccc}
\tablecaption{Effects of ISM Subtraction on Galactic Intensity Gradients 
\label{tab_gcr_grad}}
\tablehead{
\colhead{Wavelength}&
\colhead{Before/After ISM}&
\colhead{Gradient\tablenotemark{a}}&
\colhead{Correlation} \\
\colhead{($\micron$)}&
\colhead{Subtraction}&
\colhead{(nW m$^{-2}$ sr$^{-1}$ / $\csc(|b|)$)}&
\colhead{Coefficient\tablenotemark{a}}
}
\startdata
  3.5 & before & $  -4.6 \pm 0.1$ & -0.07\nl
  3.5 & after  & $  -9.2 \pm 0.1$ & -0.15\nl
  4.9 & before & $  -0.4 \pm 0.1$ & -0.005\nl
  4.9 & after  & $  -5.7 \pm 0.1$ & -0.13\nl
 12   & before & $ -36.6 \pm 0.5$ & -0.17\nl
 12   & after  & $ -70.6 \pm 0.4$ & -0.37\nl
 25   & before & $ -45.0 \pm 0.3$ & -0.33\nl
 25   & after  & $ -62.0 \pm 0.3$ & -0.45\nl
 60   & before & $  24.5 \pm 0.1$ &  0.50\nl
 60   & after  & $   1.1 \pm 0.1$ &  0.03\nl
100   & before & $  76.1 \pm 0.3$ &  0.51\nl
100   & after  & $  -3.3 \pm 0.1$ & -0.04\nl
140   & before & $ 128.7 \pm 0.6$ &  0.43\nl
140   & after  & $  34.1 \pm 0.5$ &  0.16\nl
240   & before & $  55.6 \pm 0.2$ &  0.46\nl
240   & after  & $  16.3 \pm 0.2$ &  0.22\nl
\enddata
\tablenotetext{a}{$|b|>30\arcdeg$}
\end{deluxetable}

\begin{table}
\dummytable\label{tab_errors}
\end{table}

\begin{table}
\dummytable\label{tab_pulsars}
\end{table}

\begin{deluxetable}{ccccccc}
\scriptsize
\tablecaption{Results of Fits Using $N$(H I) and $I$(H$\alpha$)
\label{tab_hi_ha}}
\tablehead{
\colhead{}&
\colhead{$A$}&
\colhead{$B$}&
\colhead{$C$}&
\colhead{$D_1$}&
\colhead{$E$}&
\colhead{}\\
\colhead{Sample}&
\colhead{($\frac{{\rm nW\ m}^{-2}\ {\rm sr}^{-1}}{10^{20}\ {\rm cm}^{-2}}$)}&
\colhead{(nW m$^{-2}$ sr$^{-1}$)}&
\colhead{($\frac{{\rm nW\ m}^{-2}\ {\rm sr}^{-1}}{10^{20}\ {\rm cm}^{-2}}$)}&
\colhead{($\frac{{\rm nW\ m}^{-2}\ {\rm sr}^{-1}}{{\rm Rayleigh}}$)}&
\colhead{(nW m$^{-2}$ sr$^{-1}$)}&
\colhead{$N_{\rm points}$}
}
\startdata
1 & $16.2\pm1.1$ & $26.9\pm6.3$ & $16.3\pm1.1$ & $-0.3\pm1.5$ & $27.4\pm8.3$ & 95 \nl
2 & $21.9\pm2.4$ & $42\pm12$    & $19.5\pm1.6$ & $10.1\pm2.3$ & $25.2\pm9.5$ & 27 \nl
3 & $23.0\pm1.4$ & $16.6\pm4.3$ & $23.1\pm1.8$ & $0.1\pm0.4$  & $16.5\pm5.4$ & 22 \nl
\enddata
\tablecomments{Fits of the forms $\nu I_{ZG_S}(100~\micron) = A\ N({\rm H\ I}) 
+ B$ and $\nu I_{ZG_S}(100~\micron) = C\ N({\rm H\ I}) + D_1\ I({\rm H}\alpha)\ 
\tau/(1-e^{\tau}) + E$.}
\end{deluxetable}

\begin{deluxetable}{ccccccc}
\scriptsize
\tablecaption{Results of Fits Using $N$(H I) and Dispersion Measure
\label{tab_hi_dm}}
\tablehead{
\colhead{}&
\colhead{$A$}&
\colhead{$B$}&
\colhead{$C$}&
\colhead{$D_2$}&
\colhead{$E$}&
\colhead{}\\
\colhead{Sample}&
\colhead{($\frac{{\rm nW\ m}^{-2}\ {\rm sr}^{-1}}{10^{20}\ {\rm cm}^{-2}}$)}&
\colhead{(nW m$^{-2}$ sr$^{-1}$)}&
\colhead{($\frac{{\rm nW\ m}^{-2}\ {\rm sr}^{-1}}{10^{20}\ {\rm cm}^{-2}}$)}&
\colhead{($\frac{{\rm nW\ m}^{-2}\ {\rm sr}^{-1}}{10^{20}\ {\rm cm}^{-2}}$)}&
\colhead{(nW m$^{-2}$ sr$^{-1}$)}&
\colhead{$N_{\rm points}$}
}
\startdata
4 & $24.3\pm2.2$ & $12.3\pm5.8$  & $21.2\pm3.7$ & $15.6\pm19.6$  & $5.9\pm9.0$ & 9 \nl
5 & $22.7\pm3.1$ & $18.9\pm12.5$ & $29.4\pm5.0$ & $-41.3\pm33.1$ & $42.9\pm20.9$ & 22 \nl
\enddata
\tablecomments{Fits of the forms $\nu I_{ZG_S}(100~\micron) = A\ N({\rm H\ I}) 
+ B$ and $\nu I_{ZG_S}(100~\micron) = C\ N({\rm H\ I}) + D_2\ DM + E$.}
\end{deluxetable}

\end{document}